%% file: main.tex
\pdfoutput=1

\newif\ifextended\extendedtrue

\newif\ifdraft\draftfalse

\let\mymarginpar\marginpar
\documentclass[a4paper,envcountsame]{llncs}

\newif\ifdraft\draftfalse

\ifextended
  \newcommand{\extendedOnly}[1]{#1}
  \newcommand{\baseOnly}[1]{}
  \newcommand{\allGraphsDir}{.}
\else
  \newcommand{\extendedOnly}[1]{}
  \newcommand{\baseOnly}[1]{#1}
\fi

\usepackage[T1]{fontenc}
\usepackage{times}

\usepackage{hyperref}

\usepackage{changepage}

\usepackage{boxedminipage}
\newenvironment{result}%
{
\noindent
\let\emph=\textbf
\begin{boxedminipage}{\columnwidth}\begin{center}\em}%
{\end{center}\end{boxedminipage}%
}

\usepackage{paralist}

\usepackage[scaled=0.81]{beramono} 
\usepackage{eiffel}
\newcommand{\eif}[1]{\mbox{\lstinline|#1|}}

\usepackage{amsmath}
\usepackage{array}
\usepackage{triotex}
\usepackage{xspace}
\usepackage[table]{xcolor}
\usepackage{graphicx}
\usepackage{lscape}
\usepackage{rotating}
\usepackage{listings}

\ifdraft
	\newcommand\rem[1]{%
          \mymarginpar{\raggedright\hbadness=10000\tiny\it #1\par}}
\else
	\newcommand\rem[1]	{}
\fi

\ifdraft
\fi

\newcommand{\fakepar}[1]{\textbf{#1}}
\newcommand{\fakeparagraph}[1]{\fakepar{#1}}

\newcommand{\comment}[2]{\rem{\textbf{{#1}}: {#2}}}
\newcommand{\caf}[1]{\comment{Carlo}{#1}}

\newcommand{\coat}{\textsc{Coat}\xspace}   

\definecolor{lightgray}{gray}{0.9}
\definecolor{shadingcolor}{rgb}{.9,.9,.9}
\newcommand{\highlightbox}{\makebox[1pt][l]{\color{shadingcolor}\rule[-4pt]{.75\linewidth}{12pt}}}

\newcommand{\mm}{\ensuremath{\mathbf{m}}}
\newcommand{\MM}{\ensuremath{\mathbf{M}}}
\newcommand{\mmedian}{\ensuremath{\mu}}
\newcommand{\std}{\ensuremath{\sigma}}

\newcommand{\figDir}{.}

\makeatletter
\renewcommand*{\thetable}{\arabic{table}}
\renewcommand*{\thefigure}{\arabic{figure}}
\let\c@table\c@figure
\makeatother

\begin{document}

\setlength{\intextsep}{7pt}
\setlength{\textfloatsep}{7pt}

\title{Contracts in Practice\thanks{Work supported by Gebert-Ruf Stiftung, by ERC grant CME \#~291389, and by SNF grant ASII~\#~200021-134976.
}}

\author{
H.-Christian Estler
\and
Carlo A.\ Furia
\and
Martin Nordio
\and\\
Marco Piccioni
\and
Bertrand Meyer
}
\institute{%
Chair of Software Engineering, Department of Computer Science, ETH Zurich, Switzerland\\
\email{firstname.lastname@inf.ethz.ch}
}

\maketitle

\begin{abstract}
Contracts are a form of lightweight formal specification embedded in the program text. Being executable parts of the code, they encourage programmers to devote proper attention to specifications, and help maintain consistency between specification and implementation as the program evolves. The present study investigates how contracts are used in the practice of software development. Based on an extensive empirical analysis of 21 contract-equipped Eiffel, C\#, and Java projects totaling more than 260 million lines of code over 7700 revisions, it explores, among other questions: 1) which kinds of contract elements (preconditions, postconditions, class invariants) are used more often; 2) how contracts evolve over time; 3) the relationship between implementation changes and contract changes; and 4) the role of inheritance in the process. It has found, among other results, that: the percentage of program elements that include contracts is above 33\% for most projects and tends to be stable over time; there is no strong preference for a certain type of contract element; contracts are quite stable compared to implementations; and inheritance does not significantly affect qualitative trends of contract usage.
\end{abstract}

\section{Introduction} \label{sec:introduction}

Using specifications as an integral part of the software development process has long been advocated by formal methods pioneers and buffs.
While today few people question the value brought by formal specifications, the software projects that systematically deploy them are still a small minority.
What can we learn from these adopters about the practical usage of specifications to support software development?

In this paper, we answer this question by looking into \emph{contracts}, a kind of light\-weight formal specification in the form of executable assertions (preconditions, postconditions, and class invariants).
In the practice of software development, contracts support a range of activities such as runtime checking, automated testing, and static verification, and provide rigorous and unambiguous API documentation.
They bring some of the advantages of ``heavyweight'' formal methods while remaining amenable to programmers without strong mathematical skills: whoever can write Boolean expressions can also write contracts.
Therefore, learning how contracts are used in the projects that use them can shed light on how formal methods can make their way into the practice of software development.

The empirical study of this paper analyzes 21 projects written in Eiffel, C\#, and Java, three major object-oriented languages supporting contracts, with the goal of studying how formal specifications are written, changed, and maintained as part of general software development. Eiffel has always supported contracts natively; the Java Modeling Language (JML~\cite{JML99}) extends Java with contracts written as comments; and C\# has recently added support with the Code Contracts framework~\cite{FahndrichBL10}. 
Overall, our study analyzed more than 260 million lines of code and specification distributed over 7700 revisions.
To our knowledge, this is the first extensive study of the practical evolving usage of simple specifications such as contracts over project lifetimes. 

The study's specific \textbf{questions} target various aspects of how contracts are used in practice: Is the usage of contracts quantitatively significant and uniform across the various selected projects? How does it evolve over time? How does it change with the overall project? What kinds of contracts are used more often? What happens to contracts when implementations change? What is the role of inheritance?

The main \textbf{findings} of the study, described in Section~\ref{sec:dataAnalysis}, include:
\begin{compactitem}
	\item The projects in our study make a \emph{significant usage} of contracts: the percentages of routines and classes with specification is above 33\% in the majority of projects.
	\item The usage of specifications tends to be \emph{stable over time}, except for the occasional turbulent phases where major refactorings are performed.
     This suggests that contracts evolve following design changes.
	\item There is \emph{no strong preference} for certain \emph{kinds} of specification elements (preconditions, postconditions, class invariants); but preconditions, when they are used, tend to be larger (have more clauses) than postconditions.
     This indicates that different specification elements are used for different purposes.
	\item Specifications are quite \emph{stable} compared to implementations: a routine's body may change often, but its contracts will change infrequently. This makes a good case for a fundamental software engineering principle: stable interfaces over changing implementations~\cite{DBLP:journals/cacm/Parnas72a}.
	\item \emph{Inheritance} does not significantly affect the qualitative findings about specification usage: measures including and excluding inherited contracts tend to correlate.
     This suggests that the abstraction levels provided by inheritance and by contracts are largely complementary.
\end{compactitem}
As a supplemental contribution, we make all data collected for the study available online as an SQL database image~\cite{COAT-data}.
This provides a treasure trove of data about practically all software projects of significant size publicly available that use contracts.

\fakepar{Positioning: what this study is \emph{not}.} 
The term ``specification'' has a broad meaning. 
To avoid misunderstandings, let us mention other practices that might be interesting to investigate, but which are \emph{not} our target in this paper.
\extendedOnly{\begin{itemize}}
\extendedOnly{\item}We do not consider formal specifications in forms other than executable contracts\extendedOnly{ (Section~\ref{sec:contracts})}.
\extendedOnly{\item}We do not look for formal specifications in \emph{generic} software projects: it is well-known~\cite{Parnas-FOSE} that the overwhelming majority of software does not come with formal specifications (or any specifications). Instead, we pick our projects among the minority of those actually using contracts, to study how the few adopters use formal specifications in practice.
\extendedOnly{\item}We do not study \emph{applications} of contracts; 
  but our analysis may serve as a basis to follow-up studies targeting applications.
\extendedOnly{\item}
We do not compare different methodologies to design and write contracts; we just observe the results of programming practices.
\extendedOnly{\end{itemize}}

\baseOnly{
\fakepar{Extended version.}
For lack of space, we can only present the most important facts; an extended version~\cite{HowSpecChange-TR-17082012} provides more details on both the analysis and the results.
}

\ifextended
\section{Contracts as Specification} \label{sec:contracts}

\ifextended
\begin{figure*}[!hbt]
\centering
\lstset{basicstyle=\scriptsize}
\begin{tabular}{m{.45\textwidth} m{.45\textwidth}}
\begin{lstlisting}
  class MEASURE  -- Revision 1

     make do create data end

     data: LIST [INTEGER] -- List of data

     add_datum (d: INTEGER)
       require not data.has (d)   (*\label{pre-add-1}*)
       do  
            data.append (d)
       ensure not data.is_empty   (*\label{post-add-1}*)

     copy_data (new_list: LIST [INTEGER])
       require new_list /=  Void    (*\label{pre-copy-1}*)
       do
           across new_list as x: add_datum (x)  

  invariant
     data /=  Void    (*\label{inv-1}*)
\end{lstlisting}
&
\begin{lstlisting}
  class MEASURE  -- Revision 2

     make do create data end

     data: LIST [INTEGER] -- List of data

     add_datum (d: INTEGER)
       require not data.has (d)     (*\label{pre-add-2}*)
(*\highlightbox \label{pre-add-2b}*)               d >= 0
       do      
           data.append (d)
       ensure  not data.is_empty   

     copy_data (new_list: LIST [INTEGER])
       require new_list /=  Void  (*\label{pre-copy-2}*)
       do
(*\highlightbox*)        if not new_list.is_empty then(*\label{if}*)
            across new_list as x: add_datum (x)   (*\label{loop}*)

  invariant
     data /=  Void    (*\label{inv-2}*)
(*\highlightbox*)     not data.is_empty  (*\label{inv-2-b}*)
\end{lstlisting}
\end{tabular}
\caption{Class \eif{MEASURE} in revision~1~(left) and 2~(right). Lines added in revision 2 are shadowed.}
\label{fig:measure}
\namelabel{fig:measure-v1}{\ref{fig:measure}~(left)}
\namelabel{fig:measure-v2}{\ref{fig:measure}~(right)}
\end{figure*}
\fi

\emph{Contracts}~\cite{Meyer97} are the form of lightweight specification that we consider in this paper; therefore, we will use the terms ``contract'' and ``specification'' as synonyms. 
\extendedOnly{This section gives a concise overview of the semantics of contracts and of how they can change with the implementation.
The presentation uses a simplified example written in pseudo-Eiffel code (see Figure~\ref{fig:measure}) which is, however, representative of features found in real code (see Section~\ref{sec:dataAnalysis}).}

\extendedOnly{Consider a class \eif{MEASURE} used to store a sequence of measures, each represented by an integer number; the sequence is stored in a list as attribute \mbox{\eif{data}.}
Figure~\ref{fig:measure} shows two revisions of class \eif{MEASURE}, formatted so as to highlight the lines of code or specification added in revision~2.
\eif{MEASURE} includes specification elements in the form of preconditions (\eif{require}), postconditions \mbox{(\eif{ensure}),} and class invariants (\eif{invariant}). 
Each element includes one or more \emph{clauses}, one per line; the clauses are logically \emph{and}ed.
For example, routine (method) \eif{add_datum} has one precondition clause on line~\ref{pre-add-1} and, in revision 2, another clause on line~\ref{pre-add-2b}.}

\baseOnly{Each contract element (precondition, postcondition, class invariant) consists of one or more \emph{clauses}, one per line; the clauses are logically \emph{and}ed.}
Contract clauses use the same syntax as Boolean expressions of the programming language; therefore, they are \emph{executable} and can be checked at runtime. 
A routine's precondition must hold whenever the routine is called; the caller is responsible for satisfying the precondition of its callee.
\baseOnly{For instance, a precondition may consists of two clauses, requiring that argument \eif{n} be positive \mbox{(\eif{n > 0})} and attribute \eif{list} be not null (\eif{list /= Void} using Eiffel syntax).}
A routine's postcondition must hold whenever the routine terminates execution; the routine body is responsible for satisfying the postcondition upon termination.
A class invariant specifies the ``stable'' object states between consecutive routine calls: it
must hold whenever a new object of the class is created and before and after every (public) routine call terminates.\footnote{The semantics of class invariants is more subtle in the general case~\cite{Meyer97} but the details are unimportant here.}
\extendedOnly{In Figure~\nref{fig:measure-v1}, routine \eif{add_datum} must be called with an actual argument \eif{d} that represents a measure not already stored in the list \eif{data} (precondition on line~\ref{pre-add-1}); when the routine terminates, the list \eif{data} must not be \eif{Void} (class invariant on line~\ref{inv-1}) and not empty (postcondition on line~\ref{post-add-1}).}

\baseOnly{Contracts may change with the code, becoming stronger, weaker, or of incomparable strength. 
Since contract clauses are logically conjoined (\emph{and}), adding a clause without changing the others makes a contract element \emph{stronger}; and removing a clause makes it \emph{weaker}.}
\extendedOnly{Revision~2 of class \eif{MEASURE}, in Figure~\nref{fig:measure-v2}, introduces changes in the code and in the specification.
Some contracts become \emph{stronger}: a precondition clause and a class invariant clause are added on lines \ref{pre-add-2b} and~\ref{inv-2-b}.
Routine \eif{copy_data} changes its implementation: revision~2 \emph{checks} whether \eif{new_list} is empty (line~\ref{if}) before copying its elements, so as to satisfy the new invariant clause on line~\ref{inv-2-b}.}

\fi

\ifextended
\section{Why We Should Care About Changing Specifications} \label{sec:overview}

Since specifications in the form of contracts are executable, their changes over the development of a software project may directly affect the ways in which the software is developed, and automatically tested and analyzed.
We now sketch a few practical examples that motivate the empirical analysis. 

\fakepar{Testing} is a widely used verification technique based on
executing a system to find failures that reveal errors.  Testing
requires \emph{oracles}~\cite{StaatsWH11,HieronsBBCDDGHKKLSVWZ09,HarmanKLMY10} to
determine if the call to a certain routine is valid and produces the
expected result.  Since contracts 
can be \emph{evaluated} at runtime like any other program
expression, they can serve as completely automatic testing oracles.
Previous work (to mention just a
few:~\cite{MarinovK01,CheonL02,BoyapatiKM02,Xie06,TillmannH08,ZimmermanN10,Meyer2009,WeiETAL11,PFPWM-ICSE13})
has built ``push button'' testing frameworks that use contracts. 

The effective usage of contracts as oracles in software
testing~\cite{Gaudel07,JiangHSZX08,AraujoBriandLabiche11} rests on some
assumptions.  Besides the obvious requirements that contracts be
available and maintained, how pre- and postconditions change affects
the testability of routines.
The stronger the precondition of a routine $r$ is, the harder is testing
the routine, because more calls to $r$ become invalid.  \extendedOnly{In
Figure~\ref{fig:measure}, \eif{add_datum} is harder to test in
revision 2 because it has a stronger precondition.}  On the other hand,
a stronger precondition makes $r$'s clients more easily testable for
errors, in that there is a higher chance that a test suite will
trigger a precondition violation that does not comply with $r$'s
stricter specification.  \extendedOnly{This is the case of \mbox{\eif{copy_data}} in
the example of Section~\ref{sec:contracts}, which calls
\eif{add_datum} and hence may fail to satisfy the latter's stronger
precondition in revision~2.}  Conversely, a stronger postcondition
makes a routine itself easier to test for errors.

\fakepar{Conflict analysis.}  
\emph{Indirect} conflicts~\cite{BrunHolmesEtAl2011,NEFM-TR-20110919} between code
maintained by different developers complicate collaborative development practices.
Specifications in the form of contracts
can help detect such conflicts: syntactic changes to the contract
of a public routine may indicate conflicts in its clients, if the
syntactic changes reflect a changed routine semantics.  Thus, using
syntactic changes as indicators of possible indirect conflicts is
workable only if contracts change much less frequently than
implementations, so that following changes in the former generates only a limited number of warnings; 
and, conversely, only if specifications are consistently
changed when the semantics of the implementation changes, so as to
produce few false negatives.

\fakepar{Object retrieval error detection.}
Changes in the attributes of a class may affect the capability to retrieve previously stored objects~\cite{MonkS93,CurinoMZ08,PiccioniOriolMeyer12}.
Class invariants help detect when inconsistent objects stored in a previous revision are introduced in the system: they express properties of the object state, and hence of attributes, that every valid object must satisfy.
\extendedOnly{In Figure~\ref{fig:measure}, objects stored in revision~1 with an empty \eif{data} list cannot be retrieved after the code has evolved into revision~2 because they break the new class invariant.
Knowing whether developers consistently add invariant clauses for describing constraints on newly introduced attributes tells us whether class invariants are reliable to detect inconsistent objects as they are retrieved.}

\fi

\section{Study Setup}\label{sec:study_setup}

\newcommand{\coatRepo}{\textsc{CoatRepo}\xspace} 
\newcommand{\coatEiffel}{\textsc{CoatEiffel}\xspace}
\newcommand{\coatC}{\textsc{CoatC\#}\xspace}  
\newcommand{\coatJava}{\textsc{CoatJava}\xspace}  
\newcommand{\coatAnalyze}{\textsc{CoatAnalyze}\xspace} 

Our study analyzes contract specifications in Eiffel, C\#, and Java, covering a wide range of projects of different sizes and life spans developed by professional programmers and researchers.
\baseOnly{We use the terms ``contract'' and ``specification'' as synonyms.}
\extendedOnly{The following subsections present how we selected the projects; the tools we developed to perform the analysis; and the raw measures collected.}

\ifextended
\subsection{Data Selection} 
\else
\fakepar{Data selection.}
\fi
\label{sec:data-selection}
We selected 21 open-source projects that use contracts and are available in public repositories. 
Save for requiring a minimal amount of revisions (at least 30) and contracts (at least 5\% of elements in the latest revisions), we included all open-source projects written in Eiffel, C\# with CodeContracts, or Java with JML  we could find when we performed this research.
Table~\ref{tab:projects} lists the projects and, for each of them, the total number of \textsc{rev}isions, the life span (\textsc{age}, in weeks), the size in lines of code (\textsc{loc}) at the latest revision, the number of \textsc{dev}elopers involved (i.e., the number of committers to the repository), and a short description.

\extendedOnly{
The 8 Eiffel projects comprise some of the largest publicly available Eiffel applications and libraries, such as the EiffelBase and Gobo libraries (maintained by Eiffel Software and GoboSoft), as well as EiffelProgramAnalysis (developed by PhD students) and AutoTest (developed by our research group).
We picked the C\# projects available on the Code Contracts webpage\extendedOnly{~\cite{codecontract-webpage}}, which lists all major C\# open projects using contracts; 4 of the C\# projects (Boogie, CCI, Dafny, and Quickgraph) were mainly developed by Microsoft Research.
The main sources of Java projects using JML were the official JML webpage\extendedOnly{~\cite{jmlspec-webpage}}, and the OpenJML\extendedOnly{~\cite{openjml-webpage}} and KindSoftware\extendedOnly{~\cite{kind-webpage}} projects.
We screened 44 projects officially using JML, but only 6 satisfied our requirements about minimal amount of contracts and revisions.

With the help of the project configuration files, we manually went through all project repositories to weed out the artifacts not part of the main application (e.g., test suites, accessory library code, or informal documentation).
When a repository contained multiple branches, we selected the main branch (\emph{trunk} in Subversion and \emph{default} in Mercurial) and excluded the others.
}

\begin{table}
\setlength{\tabcolsep}{2pt}
\begin{center}
\rowcolors{1}{}{lightgray}
\begin{scriptsize}
\begin{tabular}{|rllrrrrl|}
\hline
 \textsc{\#} & \textsc{project} & \textsc{lang.} 
 & \textsc{\# rev.} & \textsc{age} & \textsc{\# loc} & \textsc{\# dev.}  & \textsc{description} \\
\hline

1 & AutoTest & Eiffel & 306 & 195 & 65'625 & 13 & Contract-based random testing tool \\

2 & EiffelBase & Eiffel & 1342 & 1006 & 61'922& 45 & General-purpose data structures library \\

3 & EiffelProgramAnalysis & Eiffel & 208 & 114 & 40'750 & 8 & Utility library for analyzing Eiffel programs \\

4 & GoboKernel & Eiffel & 671 & 747 & 53'316 &  8 & Library for compiler interoperability \\

5 & GoboStructure & Eiffel & 282 & 716 & 21'941 & 6 & Portable data structure library \\

6 & GoboTime & Eiffel & 120 & 524 & 10'840 & 6 & Date and time library \\

7 & GoboUtility & Eiffel & 215 & 716 & 6'131 & 7 & Library to support design patterns \\
 
8 & GoboXML & Eiffel & 922 & 285 & 163'552 & 6 & XML Library supporting XSL and XPath \\

	
9 &	Boogie & C\# & 766 & 108 & 88'284 & 29 & Program verification system \\

10 & CCI & C\# & 100 & 171 & 20'602 & 3 & Library to support compilers construction  \\

11 & Dafny & C\# & 326 & 106 & 29'700 & 19 & Program verifier \\

12 & LabsFramework & C\# & 49 & 30 & 14'540 & 1 & Library to manage experiments in .NET \\
 
13 & Quickgraph & C\# & 380 & 100 & 40'820 & 4 & Generic graph data structure library \\

14 & Rxx & C\# & 148 & 68 & 55'932 & 2 & Library of unofficial reactive LINQ extensions \\

15 & Shweet & C\# & 59 & 7  & 2352 & 2 & Application for messaging in Twitter style \\


16 & DirectVCGen & Java & 376 & 119 & 13'294 & 6 & Direct Verification Condition Generator \\

17 & ESCJava & Java & 879 & 366 & 73'760 & 27 & An Extended Static Checker for Java (version 2) \\

18 & JavaFE & Java & 395 & 389 & 35'013 & 18 & Front-end parser for Java byte and source code \\

19 & Logging & Java & 29 & 106 & 5'963 & 3 & A logging framework \\

20 & RCC & Java & 30 & 350 & 10'872 & 7 & Race Condition Checker for Java \\

21 & Umbra & Java & 153 & 169 & 15'538 & 8 & Editor for Java bytecode and BML specifications \\

\hline
& \textbf{Total} & & 7'756 & 6'392 & 830'747 & 228 & \\
\hline 
\end{tabular}
\end{scriptsize}
\end{center}
\caption{List of projects used in the study. ``\textsc{age}'' is in weeks, ``\textsc{\#loc}'' is lines of code\extendedOnly{, and ``\textsc{\#Developers}'' equals the number of committers to the repository}.}\label{tab:projects}
\end{table}

\extendedOnly{
\subsection{Analysis Tools} \label{tool}
To support analysis of large amounts of program code in multiple languages, we developed \coat---a ``COntract Analysis Tool''.
The current implementation of \coat has five main components: \coatRepo retrieves the complete revision history of projects; \coatEiffel, \coatC, and \coatJava are language-specific back-ends that process Eiffel and C\# classes and extract contracts and code into a database; \coatAnalyze queries the database data supplied by the back-ends and produces the raw measures discussed in Section~\ref{measures}.
Finally, a set of R scripts read the raw data produced by \coatAnalyze and perform statistical data analysis.

\ifextended
\coatRepo accesses Subversion and Mercurial repositories, checks out all
revisions of a project, and stores them locally together with other relevant data such as commit dates, messages, and authors. 
We used this additional data to investigate unexpected behavior, such as sudden extreme changes in project sizes, as we mention in Section~\ref{sec:dataAnalysis}.

\coatEiffel parses Eiffel classes, extracts body and specification elements, and stores them in a relational database, in a form suitable for the subsequent processing.
While parsing technology is commonplace, parsing projects over a life span of nearly 20 years (such as EiffelBase) is challenging because of changes in the language syntax and semantics. 
\fi

A major question for our analysis was how to deal with \emph{inheritance}.
Routines and classes inherit contracts as well as implementations; when analyzing the specification of a routine or a class, should our measures include the inherited specification?
Since we had no preliminary evidence to prefer one approach or the other, our tools analyze each class twice: once in its \emph{flat} version and once in its \emph{non-flat} version.
The non-flat version of a class is limited to what appears in the class text.
A \textit{flat} class, in contrast, explicitly includes all the routines (with their specification) and invariants of the ancestor classes.
Flattening ignores, however, library classes or framework classes 
that are not part of the repository.

\ifextended
Reconstructing flat classes in the presence of multiple inheritance (supported and frequently used in Eiffel) has to deal with features such as member renaming and redefinitions which introduce further complexity, and hence requires static analysis of the dependencies among the abstract syntax trees of different classes.
Parsing C\# is simpler because it only deals with the latest version of the language and single inheritance.
The Code Contracts library has, however, its peculiarities that require some special processing to make the results comparable with Eiffel's.
For example, specifications of an interface or abstract class must appear in a separate child class containing only the contracts; our tool merges these ``specification classes'' with their parent.
Section~\ref{sec:inheritance} compares our measures for the flat and non-flat versions of our projects; the overall conclusion is that the measures tend to be correlated.
This is a useful piece of information for the continuation of our study: for the measures we took, both considering in detail and overlooking inheritance seem to lead to consistent results.

\coatAnalyze reads the data stored in the database by \coatEiffel, \coatC, and \coatJava and computes the raw measures described in Section~\ref{measures}.
It outputs them to CSV files, which are finally processed by a set of R scripts that produce tables with statistics (such as Table~\ref{tab:overalls-inpaper}) and plots (such as those next to Table~\ref{tab:overalls-inpaper}).
The complete set of statistics is available~\cite{COAT-data,HowSpecChange-TR-17082012}\extendedOnly{ (see the appendix)}.
\fi
}

\ifextended
\subsection{Measures} 
\else
\fakepar{Measures.}
\fi
\label{measures}
\ifextended\else To support analysis of large amounts of program code in multiple languages, we developed \coat---a ``COntract Analysis Tool''.\fi
The raw measures produced by \ifextended\coatAnalyze\else\coat\fi include: 
\extendedOnly{\begin{itemize}}
\extendedOnly{\item}the number of classes, the number of classes with invariants, the average number of invariant clauses per class, and the number of classes modified compared to the previous revision;
\extendedOnly{\item}the number of routines (public and private), the number of routines with non-empty precondition, with non-empty postcondition, and with non-empty specification (that is, precondition, postcondition, or both), the average number of pre- and postcondition clauses per routine, and the number of routines with modified body compared to the previous revision.
\extendedOnly{\end{itemize}}

Measuring precisely the \emph{strength} of a specification (which refers to how constraining it is) is hardly possible as it requires detailed knowledge of the semantics of classes and establishing undecidable properties in general\extendedOnly{ (it is tantamount to deciding entailment for a first-order logic theory)}.
In our study, we \emph{count} the number of specification clauses (elements \emph{and}ed, normally on different lines) as a proxy for specification strength. 
The number of clauses is a measure of \emph{size} that is interesting in its own right.
\extendedOnly{Then, if a (non-trivial, i.e., not identically \emph{true}) clause is added to a specification element without changing its other clauses, we certainly have a strengthening; and, conversely, a weakening when we remove a clause.}
If some clauses are changed,\footnote{We consider all concrete syntactic changes, that is all textual changes.} just counting the clauses may measure strength incorrectly.
We have evidence, however, that the error introduced by measuring strengthening in this way is small.
We manually inspected 277 changes randomly chosen,  and found 11 misclassifications (e.g., strengthening reported as weakening).
Following \cite[Eq.~5]{Martin96smallsample}, \baseOnly{this implies that,}\extendedOnly{this gives a 95\% confidence interval of $[2\%, 7\%]$:} with 95\% probability, the errors introduced by our estimate (measuring clauses for strength) involve no more than 7\% of the changes.

\section{How Contracts Are Used} \label{sec:dataAnalysis}
\extendedOnly{
This section presents the main findings of our study regarding what kinds of specifications programmers write and how they change the specifications as they change the system.}
Our study targets the following main \emph{questions}, addressed in \extendedOnly{each of}the following subsections.
\begin{enumerate}[Q1.]
\item Do projects make a significant \emph{usage} of contracts, and how does usage evolve over time?

\item How does the usage of contracts change with projects growing or shrinking in \emph{size}?

\item What \emph{kinds} of contract elements are used more often?

\item What is the typical \emph{size} and \emph{strength} of contracts, and how does it change over time?

\item Do \emph{implementations} change more often than their \emph{contracts}?

\item What is the role of \emph{inheritance} in the way contracts change over time?
\end{enumerate}
Table~\ref{tab:overalls-inpaper} shows the essential quantitative data we discuss for each project; Table~\ref{tab:eiffelbase-plots} shows sample plots of the data for four projects.
In the rest of the section, we illustrate and summarize the data in Table~\ref{tab:overalls-inpaper} and the plots in Table~\ref{tab:eiffelbase-plots} as well as much more data and plots that, for lack of space, are available elsewhere~\cite{COAT-data,HowSpecChange-TR-17082012}.

\subsection{Writing Contracts} \label{sec:writ-spec} \label{sec:using-contracts}
In the majority of projects in our study, developers devoted a considerable part of their programming effort to writing specifications for their code.
While we specifically target projects with \emph{some} specification (and ignore the majority of software that doesn't use contracts), we observe that most of the projects achieve \emph{significant} percentages of routines or classes with specification.
As shown in column \textsc{\% routines spec} of Table~\ref{tab:overalls-inpaper}, in 7 of the 21 analyzed projects, on average 50\% or more of the public routines have some specification (pre- or postcondition); in 14 projects, 35\% or more of the routines have specification; and only 3 projects have small percentages of specified routines (15\% or less).
Usage of class invariants (column \textsc{\% classes inv} in Table~\ref{tab:overalls-inpaper}) is more varied but still consistent: in 9 projects, 33\% or more of the classes have an invariant; in 10 projects, 12\% or less of the classes have an invariant.
The standard deviation of these percentages is small for 11 of the 21 projects, compared to the average value over all revisions: the latter is at least five times larger. suggesting that deviations from the average are normally small. 
Section~\ref{sec:spec-proj-size} gives a quantitative confirmation of this hint about the stability of specification amount over time. 

The EiffelBase project---a large standard library used in most Eiffel projects---is a good ``average'' example of how contracts may materialize over a project's lifetime.
After an initial fast growing phase (see the first plot in Table~\ref{tab:eiffelbase-plots}), corresponding to a still incipient design that is taking shape, the percentages of routines and classes with specification stabilize around the median values with some fluctuations that---while still significant, as we comment on later---do not affect the overall trend or the average percentage of specified elements.
This two-phase development (initial mutability followed by stability) is present in several other projects of comparable size, and is sometimes extreme, such as for Boogie, where there is a widely varying initial phase, followed by a very stable one where the percentages of elements with specification is practically constant around 30\%.
Analyzing the commit logs around the revisions of greater instability showed that wild variations in the specified elements coincide with major reengineering efforts.
For Boogie, the initial project phase coincides with the porting of a parent project written in Spec\# (a dialect of C\#), and includes frequent alternations of adding and removing code from the repository; after this phase, the percentage of routines and classes with specification stabilizes to a value close to the median.

There are few outlier projects where the percentage of elements with specification is small, not kept consistent throughout the project's life, or both.
Quickgraph, for example, never has more than 4\% of classes with an invariant or routines with a postcondition, and its percentage of routines with precondition varies twice between 12\% and 21\% in about 100 revisions (see complete data in~\cite{HowSpecChange-TR-17082012}).

\begin{result}
In two thirds of the projects, on average $1/3$ or more of the routines \\ have \emph{some} specification (pre- or postconditions).
\end{result}

\fakepar{Public vs.\ private routines.}
The data analysis focuses on contracts of \emph{public} routines.
To determine whether trends are different for \emph{private} routines, we visually inspected the plots~\cite{COAT-data} and computed the correlation coefficient\footnote{All correlation measures in the paper employ Kendall's rank correlation coefficient $\tau$.} $\tau$ for the evolution of the percentages of specified public routines against those of private routines.
The results suggest to partition the projects into three categories.
For the 9 projects in the first category---AutoTest, EiffelBase, Boogie, CCI, Dafny, JavaFE, Logging, RCC and Umbra---the correlation is positive ($ 0.51 \leq \tau \leq 0.94$) and highly significant.
The 2 projects in the second category---GoboStructure and Labs---have negative ($\tau \leq -0.47$) and also significant correlation.
The remaining 10 projects belong to the third category, characterized by correlations small in absolute value, positive or negative, or statistically insignificant.
This partitioning seems to correspond to different approaches to interface design and encapsulation: for projects in the first category, public and private routines always receive the same amount of specification throughout the project's life; projects in the second category show negative correlations that may correspond to changes to the visibility status of a significant fraction of the routines; visual inspection of projects in the third category still suggests positive correlations between public and private routines with specification, but the occasional redesign upheaval reduces the overall value of $\tau$ or the confidence level.
In fact, the confidence level is typically small for projects in the third category; and it is not significant ($p = 0.418$) only for EiffelProgramAnalysis which also belongs to the third category.
Projects with small correlations tend to be smaller in \emph{size} with fewer routines and classes; conversely, large projects may require a stricter discipline in defining and specifying the interface and its relations with the private parts, and have to adopt consistent approaches throughout their lives.

\begin{result}
In roughly half of the projects, the amounts of contracts in \emph{public} and in \emph{private} routine correlate; in the other half, correlation vanishes due to redesign changes.
\end{result}

\subsection{Contracts and Project Size} \label{sec:spec-proj-size}

\extendedOnly{In Section~\ref{sec:writ-spec}, we observed that the percentage of specified routines and classes is fairly stable over time, especially for large projects in their maturity.
We analyzed the correlation between measures of elements with specification and project size, and corroborated the found correlations with visual inspection of the graphs.}

The correlation between the number of routines or classes with some specification and the total number of routines or classes (with or without specification) is consistently strong and highly significant.
Looking at routines, 10 projects exhibit an almost perfect correlation with $\tau > 0.9$ and $p \sim 0$; only 3 projects show medium/low correlations (Labs and Quickgraph with $\tau = 0.48$, and Logging with $\tau = 0.32$) which are however still significant.
The outlook for classes is quite similar: the correlation between number of classes with invariants and number of all classes tends to be high. 
Outliers are the projects Boogie and JavaFE with the smaller correlations $\tau = 0.28$ and $\tau = 0.2$, but visual inspection still suggests that a sizable correlation exists for Boogie (the results for JavaFE are immaterial since it has only few invariants overall).
In all, the absolute number of elements with specification is normally synchronized to the overall size of a project, confirming the suggestion of Section~\ref{sec:writ-spec} that the percentage of routines and classes with specification is \emph{stable} over time.

Having established that, in general, specification and project size have similar trends, we can look into finer-grained variations of specifications over time. 
To estimate the \emph{relative} effort of writing specifications, we measured the correlation between \emph{percentage} of specified routines or classes and \emph{number} of all routines or all classes.

A first large group of projects, almost half of the total whether we look at routines or classes, show weak or negligible correlations ($-0.35 < \tau < 0.35$).
In this majority of projects, the relative effort of writing and maintaining specifications evolves largely independently of the project size.
Given that the overall trend is towards stable percentages, the high variance often originates from initial stages of the projects when there were few routines or classes in the system and changes can be momentous.
Gobo\-Kernel and DirectVCGen are specimens of these cases: the percentage of routines with contracts varies wildly in the first 100 revisions when the system is still small and the developers are exploring different design choices and styles.

Another group of 3 projects (AutoTest, Boogie, and Dafny) show strong \emph{negative} correlations ($\tau < -0.75$) both between percentage of specified routines and number of routines and between percentage of specified classes and number of classes.
The usual cross-inspection of plots and commit logs points to two independent phenomena that account for the negative correlations.
The first is the presence of large merges of project branches into the main branch; these give rise to strong irregularities in the absolute and relative amount of specification used, and may reverse or introduce new specification styles and policies that affect the overall trends.
As evident in the second plot of Table~\ref{tab:eiffelbase-plots}, AutoTest epitomizes this phenomenon, with its history clearly partitioned into two parts separated by a large merge at revision 150.
\extendedOnly{Before the merge, the system is smaller with high percentages of routines and classes with specification; with the merge, the system grows manifold and continues growing afterward, while the percentage of elements with specification decreases abruptly and then (mostly for class invariants) continues decreasing.}
The second phenomenon that may account for negative correlations \extendedOnly{between percentage of specified elements and measures of project size} is a sort of ``specification fatigue'' that kicks in as a project becomes mature and quite large.
At that point, there might be diminishing returns for supplying more specification, and so the percentage of elements with specification gracefully decreases while the project grows in size.
(This is consistent with Schiller et al.'s suggestion~\cite{writingcontracts} that annotation burden limits the extent to which contracts are used.)
The fatigue is, however, of small magnitude if present at all, and may be just be a sign of reached maturity where a solid initial design with plenty of specification elements pays off in the long run to the point that less relative investment is sufficient to maintain a stable level of maintainability and quality.

The remaining projects have significant \emph{positive} correlations ($\tau > 0.5$) between either percentage of specified routines and number of routines or between percentage of specified classes and number of classes, but not both.
In these special cases, it looks as if the fraction of programming effort devoted to writing specification tends to increase with the absolute size of the system: when the system grows, proportionally more routines or classes get a specification.
However, visual inspection suggests that, in all cases, the trend is ephemeral or contingent on transient phases where the project size changes significantly in little time.
As the projects mature and their sizes stabilize, the other two trends (no correlation or negative correlation) emerge in all cases.

\begin{result}
The fraction of routines and classes with some specification is quite \emph{stable} over time. 
Local exceptions are possible when major redesign changes take place.
\end{result}

\subsection{Kinds of Contract Elements} \label{sec:kinds-spec-elem}
Do programmers prefer preconditions?
Typically, one would expect that preconditions are simpler to write than postconditions (and, for that matter, class invariants): postconditions are predicates that may involve two states (before and after routine execution).
Furthermore, programmers have immediate benefits in writing preconditions as opposed to postconditions: a routine's precondition defines the valid input; hence, the stronger it is, the fewer cases the routine's body has to deal with.

Contrary to this common assumption, the data in our study (columns \textsc{\% routines pre} and \textsc{post} in Table~\ref{tab:overalls-inpaper}) is not consistently lopsided towards preconditions.
2 projects show no difference in the median percentages of routines with precondition and with postcondition.
10 projects do have, on average, more routines with precondition than routines with postcondition, but the difference in percentage is less than 10\% in 5 of those projects, and as high as 39\% only in one project (Dafny).
The remaining 9 projects even have more routines with postcondition than routines with precondition, although the difference is small (less than 5\%) in 5 projects, and as high as 45\% only in RCC.

On the other hand, in 17 projects the percentage of routines with some specification (precondition, postcondition, or both) is higher than both percentages of routines with precondition and of routines with postcondition.
Thus, we can partition the routines of most projects in three groups of comparable size: routines with only precondition, routines with only postcondition, and routines with both.
The 4 exceptions are CCI, Shweet, DirectVCGen, and Umbra where, however, most elements have little specification.
In summary, many exogenous causes may concur to determine the ultimate reasons behind picking one kind of contract element over another, such as the project domain and the different usage of different specification elements.
Our data is, however, consistent with the notion that programmers choose which specification to write according to context and requirements, not based on a priori preferences.
It is also consistent with Schiller et al.'s observations~\cite{writingcontracts} that contract usage follows different patterns in different projects, and that programmers are reluctant to change their preferred usage patterns---and hence patterns tend to remain consistent within the same project.

A closer look at the projects where the difference between percentages of routines with precondition and with postcondition is significant (9\% or higher) reveals another interesting pattern.
All 6 projects that favor preconditions are written in C\# or Java: Dafny, Labs, Quickgraph, Shweet, ESCJava (third plot in Table~\ref{tab:eiffelbase-plots}, after rev.~400), and JavaFE; conversely, the 3 of 4 projects that favor postconditions are in Eiffel (AutoTest, GoboKernel, and GoboTime), whereas the fourth is RCC written in Java.
A possible explanation for this division involves the longer time that Eiffel has supported contracts and the principal role attributed to Design by Contract within the Eiffel community.
\extendedOnly{C\# and Java programmers, then, are more likely to pragmatically go for the immediate tangible benefits brought by preconditions as opposed to postconditions; Eiffel programmers might be more zealous and use contracts thoroughly also for design before implementation.}

\begin{result}
Preconditions and postconditions are used \emph{equally frequently} across most projects.
\end{result}

\fakepar{Class invariants.}
Class invariants have a somewhat different status than pre- or postconditions.
Since class invariants must hold between consecutive routine calls, they define object consistence, and hence they belong to a different category than pre- and postconditions. 
The percentages of classes with invariant (\textsc{\% classes inv} in Table~\ref{tab:overalls-inpaper}) follow similar trends as pre- and postconditions in most projects in our study.
Only 4 projects stick out because they have 4\% or less of classes with invariant, but otherwise make a significant usage of other specification elements: Quickgraph, EiffelProgramAnalysis, Shweet, and DirectVCGen.\footnote{While the projects CCI and Umbra have few classes with invariants (4\%--6\%), we don't discuss them here because they also only have few routines with preconditions or postconditions.}
Compared to the others, Shweet has a short history and EiffelProgramAnalysis involves students as main developers rather than professionals.
Given that the semantics of class invariants is less straightforward than that of pre- and postconditions---and can become quite intricate for complex programs~\cite{DBLP:journals/cacm/BarnettFLMSV11}---this might be a factor explaining the different status of class invariants in these projects.
A specific design style is also likely to influence the usage of class invariants, as we further comment on in Section~\ref{sec:spec-strength}.

\fakepar{Kinds of constructs.}
An additional classification of contracts is according to the constructs they use. 
We gathered data about constructs of three types: expressions involving checks that a reference is \lstinline|Void| (Eiffel) or \textbf{null} (C\# and Java); some form of finite quantification (constructs for $\forall/\exists$ over containers exist for all three languages); and \lstinline|old| expressions (used in postconditions to refer to values in the pre-state).
\mbox{\lstinline|Void|/\textbf{null}} checks are by far the most used: in Eiffel, 36\%--93\% of preconditions,  7\%--62\% of postconditions, and 14\%--86\% of class invariants include a \lstinline|Void| check; in C\#, 80\%--96\% of preconditions contain \mbox{\textbf{null}} checks, as do 34\%--92\% of postconditions (the only exception is CCI which does not use postconditions) and 97\%--100\% of invariants (exceptions are Quickgraph at 20\% and Shweet which does not use invariants); in Java, 88\%--100\% of preconditions, 28\%--100\% of postconditions, and 50\%--77\% of class invariants contain \mbox{\textbf{null}} (with the exception of Umbra which has few contracts in general).
\lstinline|Void|/\textbf{null} checks are simple to write, and hence cost-effective, which explains their wide usage; this may change in the future, with the increasing adoption of static analyses which supersede such checks \cite{bertrand-void-safe-eiffel,DietlDEMS11}.
The predominance of simple contracts and its justification have been confirmed by others~\cite{writingcontracts}.

At the other extreme, quantifications are very rarely used: practically never in pre- or postconditions; and very sparsely (1\%--10\% of invariants) only in AutoTest, Boogie, Quickgraph, ESCJava, and JavaFE's class invariants.
This may also change in the future, thanks to the progresses in inferring complex contracts~\cite{HenkelRD07,WeiFKM11,WasylkowskiZ11}, and in methodological support~\cite{PFPWM-ICSE13}.

The usage of \lstinline|old| is more varied: C\# postconditions practically don't use it, Java projects rarely use it (2\%--3\% of postconditions at most),  whereas it features in as many as 39\% of postconditions for some Eiffel projects.
Using \lstinline|old| may depend on the design style; for example, if most routines are side-effect free and return a value function solely of the input arguments there is no need to use \lstinline|old|.

\begin{result}
The overwhelming majority of contracts involves \lstinline|Void|/\textbf{null} checks. \\
In contrast, quantifiers appear very rarely in contracts.
\end{result}

\subsection{Contract Size and Strength} \label{sec:spec-strength}
The data about specification \emph{size} (and strength) partly vindicates the intuition that preconditions are more used.
While Section~\ref{sec:kinds-spec-elem} showed that routines are not more likely to have preconditions than postconditions, preconditions have more clauses on average than postconditions in all but the 3 projects GoboTime, ESCJava, and Logging.
As shown in columns \textsc{avg routines pre} and \textsc{post} of Table~\ref{tab:overalls-inpaper}, the difference in favor of preconditions is larger than 0.5 clauses in 9 projects, and larger than 1 clause in 3 projects.
CCI never deploys postconditions, and hence its difference between pre- and postcondition clauses is immaterial.
GoboTime is a remarkable outlier: not only do twice as many of its routines have a postcondition than have precondition, but its average postcondition has 0.66 more clauses than its average precondition. 
ESCJava and Logging also have larger postconditions on average but the size difference is less conspicuous (0.25 and 0.32 clauses). 
We found no simple explanation for these exceptions, but they certainly are the result of deliberate design choices.

The following two facts corroborate the idea that programmers tend to do a better job with preconditions than with postconditions---even if they have no general preference for one or another.
First, the default ``trivial'' precondition \emph{true} is a perfectly reasonable precondition for routines that compute total functions---defined for every value of the input; a trivial postcondition is, in contrast, never satisfactory.
Second, in general, ``strong'' postconditions are more complex than ``strong'' preconditions~\cite{PFPWM-ICSE13} since they have to describe more complex relations.


Class invariants are not directly comparable to pre- and postconditions, and their usage largely depends on the design style.
Class invariants apply to all routines and attributes of a class, and hence they may be used extensively and involve many clauses; conversely, they can also be replaced by pre- and postconditions in most cases, in which case they need not be complex or present at all\extendedOnly{~\cite{Parkinson2007}}.
In the majority of projects (15 out of 21), however, class invariants have more clauses on average than pre- and postconditions. 
We might impute this difference to the traditional design principles for object-oriented contract-based  programming, which attribute a significant role to class invariants~\cite{Meyer97,DrossopoulouFMS08,semicola} as the preferred way to define valid object state.

\begin{result}
In over eighty percent of the projects, the average \emph{preconditions} \\ contain \emph{more clauses} than the average postconditions.
\end{result}

Section~\ref{sec:writ-spec} observed the prevailing stability over time of routines with specification.
Visual inspection and the values of standard deviation point to a qualitatively similar trend for specification size, measured in number of clauses.
In the first revisions of a project, it is common to have more varied behavior, corresponding to the system design being defined; but the average strength of specifications typically reaches a plateau, or varies quite slowly, in mature phases.

Project Labs is somewhat of an outlier, where the evolution of specification strength over time has a rugged behavior (see~\cite{HowSpecChange-TR-17082012} for details and plots).
Its average number of class invariant clauses has a step at about revision 29, which corresponds to a merge, when it suddenly grows from 1.8 to 2.4 clauses per class.
During the few following revisions, however, this figure drops quickly until it reaches a value only slightly higher than what it was before revision 29.
What probably happened is that the merge mixed classes developed independently with different programming styles (and, in particular, different attitudes towards the usage of class invariants).
Shortly after the merge, the developers refactored the new components to make them comply with the overall style, which is characterized by a certain average invariant strength.

One final, qualitative, piece of data about specification strength
is that in a few projects there seems to be a moderate increase in the strength of postconditions towards the latest revisions of the project.
\extendedOnly{
If this is a real phenomenon, it may show that, as programmers become fluent in writing specification, they are confident enough to go for the more complex postconditions, reversing their initial (moderate) focus on preconditions.
}
This observation is however not applicable to any of the largest and most mature projects we analyzed (e.g., EiffelBase, Boogie, Dafny).

\begin{result}
The average \emph{size} (in number of clauses) of specification elements is stable over time.
\end{result}

\newcommand{\printNonFlatSpecs}[4]{
\begin{figure}[!htb]
\begin{adjustwidth}{-2cm}{-2cm}
\begin{center}
    \begin{tabular}{cccc}
\extendedOnly{%
	\includegraphics[width=5.0cm,height=3.8cm]{\figDir/AutoTest_72--#1.pdf} &
	\includegraphics[width=5.0cm,height=3.8cm]{\figDir/AutoTest_72--#2.pdf} &
	\includegraphics[width=5.0cm,height=3.8cm]{\figDir/AutoTest_72--#3.pdf} \\
}
	\includegraphics[width=5.0cm,height=3.8cm]{\figDir/Boogie_10--#1.pdf} &
	\includegraphics[width=5.0cm,height=3.8cm]{\figDir/Boogie_10--#2.pdf} &
	\includegraphics[width=5.0cm,height=3.8cm]{\figDir/Boogie_10--#3.pdf} \\
	
	\includegraphics[width=5.0cm,height=3.8cm]{\figDir/EiffelBase_53--#1.pdf} &
	\includegraphics[width=5.0cm,height=3.8cm]{\figDir/EiffelBase_53--#2.pdf} &
	\includegraphics[width=5.0cm,height=3.8cm]{\figDir/EiffelBase_53--#3.pdf} \\
\extendedOnly{%
	\includegraphics[width=5.0cm,height=3.8cm]{\figDir/GoboKernel_70--#1.pdf} &
	\includegraphics[width=5.0cm,height=3.8cm]{\figDir/GoboKernel_70--#2.pdf} &
	\includegraphics[width=5.0cm,height=3.8cm]{\figDir/GoboKernel_70--#3.pdf} \\

	\includegraphics[width=5.0cm,height=3.8cm]{\figDir/Labs_5--#1.pdf} &
	\includegraphics[width=5.0cm,height=3.8cm]{\figDir/Labs_5--#2.pdf} &
	\includegraphics[width=5.0cm,height=3.8cm]{\figDir/Labs_5--#3.pdf} \\
}	
\extendedOnly{%
   \includegraphics[width=5.0cm,height=3.8cm]{\figDir/Quickgraph_18--#1.pdf} &
	\includegraphics[width=5.0cm,height=3.8cm]{\figDir/Quickgraph_18--#2.pdf} &
	\includegraphics[width=5.0cm,height=3.8cm]{\figDir/Quickgraph_18--#3.pdf}
}
 \end{tabular}	
\end{center}
\end{adjustwidth}
\caption{#4}\label{tab:specs}
\end{figure}
}


\subsection{Implementation vs.\ Specification Changes} \label{sec:impl-vs-spec}

\extendedOnly{
A phenomenon commonly attributed~\cite{Parnas-FOSE} to specifications is that they are not updated to reflect changes in the implementation: even when programmers deploy specifications extensively in the initial phases of a project, the later become obsolete and, eventually, useless. 
Is there evidence of such a phenomenon in the data of our projects?

We first have to remember a peculiarity of contracts as opposed to other forms of specification.}
Contracts are \emph{executable} specifications; normally, they are checked at runtime during debugging and regression testing sessions (and possibly also in production releases, if the overhead is acceptable, to allow for better error reporting from final users).
Specifically, most applications and libraries of our study are actively used and maintained. %
Therefore, their contracts cannot become grossly misaligned with the implementation\extendedOnly{: inconsistencies quickly generate runtime errors, which can only be fixed by reconciling implementations with their specifications}.
\extendedOnly{
By and large, the fact that a significant percentage of routines and classes in our study have contracts (Section~\ref{sec:writ-spec}) implies that most of them are correct---if incomplete---specifications of routine or class behavior.
}

A natural follow-up question is then whether contracts change more often or less often than the implementations they specify.
To answer, we compare two measures in the projects: for each revision, we count the number of routines with changed body and changed specification (pre- or postcondition) and compare it to the number of routines with changed body and unchanged specification.
These measures aggregated over all revisions determine a pair of values $(c_P, u_P)$ for each project $P$: $c_P$ characterizes the frequency of changes to implementations that also caused a change in the contracts, whereas $u_P$ characterizes the frequencies of changes to implementations only.
To avoid that few revisions with very many changes dominate the aggregate values for a project, each revision contributes with a binary value to the aggregate value of a project: 0 if no routine has undergone a change of that type in that revision, and 1 otherwise.\footnote{Using other ``reasonable'' aggregation functions (including exact counting) leads to qualitatively similar results.}
We performed a Wilcoxon signed-rank test comparing the $c_P$'s to the $u_P$'s across all projects to determine if the median difference between the two types of events (changed body with and without changed specification) is statistically significant.
The results confirm with high statistical significance ($V = 0$, $p = 9.54 \cdot 10^{-7}$, and large effect size---Cohen's $d > 0.99$) that specification changes are quite infrequent compared to implementation changes for the same routine.
Visual inspection also confirms the same trend: see the last plot in Table~\ref{tab:eiffelbase-plots} about Boogie.
A similar analysis ignoring routines with trivial (empty) specification leads to the same conclusion also with statistical significance ($V = 29$, $p = 4.78 \cdot 10^{-3}$, and medium effect size $d > 0.5$).

When specifications do change, what happens to their \emph{strength} measured in number of clauses?
Another Wilcoxon signed-rank test compares the changes to pre- and postconditions and class invariants that added clauses (suggesting strengthening) against those that removed clauses (suggesting weakening).
Since changes to specifications are in general infrequent, the results were not as conclusive as those comparing specification and implementation changes.
The data consistently points towards strengthening being more frequent than weakening: $V = 31.5$ and $p < 0.02$ for precondition changes; $V = 29$ and $p < 0.015$ for postcondition changes; $V = 58.5$ and $p = 0.18$ for invariant changes.
The effect sizes are, however, smallish: Cohen's $d$ is about $0.4$, $0.42$, and $0.18$ for preconditions, postconditions, and invariants.
In all, the effect of strengthening being more frequent than weakening seems to be real but more data is needed to obtain conclusive evidence.

\begin{result}
The \emph{implementation} of an average routine changes \\much more frequently than its \emph{specification}.
\end{result}

\subsection{Inheritance and Contracts} \label{sec:inheritance}

Inheritance is a principal feature of object-oriented programming, and involves contracts as well as implementations; we now evaluate its effects on the findings previously discussed.

We visually inspected the plots and computed correlation coefficients for the percentages and average strength of specified elements 
in the flat (explicitly including all routines and specification of the ancestor classes) and non-flat (limited to what appears in the class text) versions of the classes.
In the overwhelming majority of cases, the correlations are high and statistically significant: 16 projects have $\tau \geq 0.54$ and $p < 10^{-9}$ for the percentage of routines with specification; 17 projects have $\tau \geq 0.66$ and $p \sim 0$ for the percentage of classes with invariant; 12 projects have $\tau \geq 0.58$ and $p < 10^{-7}$ for the average precondition and postcondition strength (and 7 more projects still have $\tau \geq 0.33$ and visually evident correlations); and 15 projects have $\tau \geq 0.45$ and $p \sim 0$ for the average invariant strength.
The first-order conclusion is that, in most cases, ignoring the inherited specification does not preclude understanding qualitative trends.

What about the remaining projects, which have small or insignificant correlations for some of the measures in the flat and non-flat versions?
Visual inspection often confirms the absence of significant correlations, in that the measures evolve along manifestly different shapes in the flat or non-flat versions; the divergence in trends is typically apparent in the revisions where the system size changes significantly, where the overall design---and the inheritance hierarchy---is most likely to change.
To see if these visible differences invalidate some of the findings discussed so far, we reviewed the findings against the data for \emph{flat} classes.
The big picture was not affected: considering inheritance may affect the measures and offset or bias some trends, but the new measures are still consistent with the same conclusions drawn from the data for non-flat classes.
Future work will investigate whether this result is indicative of a mismatch between the semantics of inheritance and how it is used in practice~\cite{TemperoYN13,PradelG13}.
\ifextended
We now discuss the various questions for flat classes in more detail.
\else
(See the extended version~\cite{HowSpecChange-TR-17082012} for details.)
\fi

\begin{result}
Qualitative trends of measures involving contracts do \emph{not} change significantly \\ whether we consider or ignore \emph{inherited} contracts.
\end{result}

\extendedOnly{
\textbf{Usage (and kinds)}
of contracts is qualitatively similar for the flat and non-flat classes. Of course, the number of elements with specification tends to be larger in flat classes simply because specifications are inherited.
However, the \emph{relative} magnitude of measures such as the average, minimum, maximum, and standard deviation of routines and classes with specification is quite similar for flat and non-flat.
Similar observations hold concerning measures of contract \textbf{strength} and their evolution.

\textbf{Project size}
is correlated to measures regarding contracts in similar ways in flat and non-flat classes. The few outliers are projects that exhibited a positive correlation between percentage of specified elements and number of elements in the non-flat versions (e.g., $\tau = 0.66$ for EiffelBase); the correlation vanishes in the flat versions (e.g., $\tau = -0.08$ for EiffelBase).
As we discussed at the end of Section~\ref{sec:spec-proj-size}, the positive correlations of these outliers were unusual and possibly ephemeral; the fact that correlations dilute away when we consider the inheritance hierarchy reinforces the idea that the positive correlation trends of Section~\ref{sec:spec-proj-size} are exceptional and largely inconsequential.
Java projects are different in that they achieve consistent positive correlations between percentage of specified elements and number of elements in both flat and non-flat versions; there is a simple explanation for this: inheritance hierarchies are shallow in Java projects, and hence inheritance is simply negligible.
\caf{Check it is true.}

\textbf{Change}
analyses are virtually identical in flat and non-flat classes; this is unsurprising since the analyses (discussed in Section~\ref{sec:impl-vs-spec}) target binary \emph{differences} between a version and the next one, so that the measures gobble up the offset introduced by flattening.

Differences between measures with flat and non-flat classes tend to be smaller in C\# and Java projects, as opposed to the Eiffel projects.
This reveals that \emph{multiple} inheritance, available in Eiffel but not in C\# and Java, may contribute to magnify differences between measures of flat and non-flat classes.
(This is also consistent with the observation about shallow inheritance hierarchies in most Java project.)
}

\section{Threats to Validity}\label{sec:threatsToValidity}

\fakeparagraph{Construct validity.}
\extendedOnly{The measures taken by \coat expose two potential threats to construct validity.}
\extendedOnly{First, u}\baseOnly{U}sing the number of clauses as a proxy for the strength of a specification may produce imprecise measures; Section~\ref{measures}, however, estimated the imprecision and showed it is limited, and hence an acceptable trade-off in most cases (also given that computing strength exactly is infeasible).
Besides, the number of clauses is still a valuable size/complexity measure in its own right (Section~\ref{sec:spec-strength}).
\extendedOnly{
Second, the flattening introduced to study the effect of inheritance \extendedOnly{(Section~\ref{tool})} introduces some approximations when dealing with the most complex usages of multiple inheritance (\eif{select} clauses) or of inner classes.
We are confident, however, that this approximation has a negligible impact on our measurements as these complex usages occur very rarely.
}

\fakeparagraph{Internal validity.}
Since we targeted object-oriented languages where inheritance is used pervasively, it is essential that the inheritance structure be taken into account in the measures.
We fully addressed this major threat to internal validity by analyzing all projects twice: in non-flat and flat version (Section~\ref{sec:inheritance}).
\extendedOnly{
A different threat originates from \coat failing to parse a few files in some revisions, due to the presence of invalid and outdated language syntax constructs.
The impact of this is certainly negligible: less than 0.0069\% of all files could not be parsed.
Restricting our analysis to the main branches and manually discarding irrelevant content from the repositories pose another potential threat to internal validity.
In all cases, we took great care to cover the most prominent development path and to select the main content based on the project configuration files written by the developers, so as to minimize this threat.
}

\fakeparagraph{External validity.}
Our study is restricted to three formalisms for writing contract specifications\extendedOnly{: Eiffel, C\# with Code Contracts, and Java with JML}.
While other notations for contracts \extendedOnly{(e.g., other Java contract libraries or SPARK Ada)} are similar, we did not analyze other types of formal specification, which might limit the generalizability of our findings.
In contrast, the restriction to open-source projects does not pose a serious threat to external validity in our study, because several of our projects are mainly maintained by professional programmers (EiffelBase and Gobo projects) or by professional researchers in industry (Boogie, CCI, Dafny, and Quickgraph). 

%
%

An important issue to warrant external validity involves the \emph{selection of projects.}
We explicitly targeted projects that make a non-negligible usage of contracts (Section~\ref{sec:data-selection}), as opposed to the overwhelming majority that only include informal documentation or no documentation at all.
This deliberate choice limits the generalizability of our findings, but also focuses the study on understanding how contracts can be seriously used in practice. 
A related observation is that the developers of several of the study's projects are supporters of using formal specifications.
While this is a possible source of bias it also contributes to reliability of the results: since we are analyzing good practices and success stories of writing contracts, we should target competent programmers with sufficient experience, rather than inexpert novices.
\extendedOnly{For the same reason, the most extensive studies on the usage of design patterns \extendedOnly{(e.g.~\cite{DBLP:journals/smr/RiccaPTTC07}) }
target software such as Eclipse---written by experts on design patterns---rather than small student projects or low-quality software.}
Besides, Schiller et al.'s independent analysis~\cite{writingcontracts} of some C\# projects using CodeContracts also included in our study suggests that their developers are hardly fanatic about formal methods, as they use contracts only to the extent that it remains inexpensive and cost-effective, and does not require them to change their programming practices.

Nevertheless, to get an idea of whether the programmers we studied really have incomparable skills, we also set up a small \emph{control group}, consisting of 10 projects developed by students of a software engineering course\extendedOnly{\footnote{\url{http://se.inf.ethz.ch/research/dose}}} involving students from universities all around the world.
\extendedOnly{
The students use Eiffel but usually have limited or no experience with it when starting the course; the selected projects have a development time of circa 6 weeks and numbers of revisions range between 30 to 292. The average size per project is 4185 LOC. All projects were graded \emph{good} or \emph{very good} in their final evaluations.
We ran the analyses described in this paper on these student projects.}
In summary\baseOnly{ (see~\cite{HowSpecChange-TR-17082012} for details)}\extendedOnly{ (see Section~\ref{sec:control-group-dose} in the Appendix for detailed data)}, 
we found that several of the trends measured with the professional programmers were also present in the student projects---albeit on the smaller scale of a course project.
This gives some confidence that the big picture outlined by this paper's results somewhat generalizes to developers willing to spend some programming effort to write contracts\extendedOnly{ and is not limited to ``verification wonks'' only}.

\section{Related Work} \label{sec:earlierWork}
\extendedOnly{Section~\ref{sec:overview} discussed program analysis and other activities where contracts can be useful.}
To our knowledge, this paper is the first quantitative empirical study of specifications in the form of contracts and their evolution together with code.
Schiller et al.~\cite{writingcontracts} study C\# projects using CodeContracts (some also part of our study); while our and their results are not directly comparable because we take different measures and classify contract usage differently, the overall qualitative pictures are consistent and nicely complementary.
In the paper we also highlighted a few points where their results confirm or justify ours.
Schiller et al.\ do not study contract \emph{evolution}; there is evidence, however, that other forms of documentation---e.g., comments~\cite{FluriWuerschGall07}, APIs~\cite{DBLP:conf/icse/KimCK11}, or tests~\cite{ZaidmanVanEtAl2008}---evolve with code.

A well-known problem is that specification and implementation tend to diverge over time; this is more likely for documents such as requirements and architectural designs that are typically developed and stored separately from the source code.
Much research has targeted this problem; specification refinement, for instance, can be applied to software revisions~\cite{Duque09}.
Along the same lines, some empirical studies analyzed how requirements relate to the corresponding implementations; \cite{HindleBirdZimmermannNagappan12}, for example, examines the co-evolution of certain aspects of requirements documents with change logs and shows that topic-based requirements traceability can be automatically implemented from the information stored in version control systems.

The information about the usage of formal specification by programmers is largely anecdotal, with the exceptions of a few surveys on industrial practices~\cite{Chalin06,Woodcock-survey}.
There is, however, some evidence of the usefulness of contracts and assertions. \cite{KudrjavetsNB06}, for example, suggests that increases of assertions density and decreases of fault density correlate.
\cite{MullerTH02} reports that using assertions may decrease the effort necessary for extending existing programs and increase their reliability.
In addition, there is evidence that developers are more likely to use contracts in languages that support them natively~\cite{Chalin06}.
As the technology to infer contracts from code reaches high precision levels~\cite{ErnstPGMPTX07,WeiFKM11}, it is natural to compare automatically inferred and programmer-written contracts; they turn out to be, in general, different but with significant overlapping~\cite{PolikarpovaCM09}.

\extendedOnly{
Our \coat tool (Section~\ref{tool}) is part of a very large family of tools~\cite{GermanCubranicStorey05} 
that mine software repositories to extract quantitative data.
In particular, it shares some standard technologies with other tools for source code analysis (e.g.,~\cite{NeamtiuFosterHicks05}).
}

\section{Concluding Discussion \& Implications of the Results} \label{sec:discussion}
Looking at the big picture, our empirical study suggests a few actionable remarks.
\begin{inparaenum}[(\itshape i)]
\item
The effort required to make a quantitatively significant usage of lightweight specifications is sustainable consistently over the lifetime of software projects.
This supports the practical applicability of methods and processes that rely on \emph{some} form of rigorous specification.
\item
The overwhelming majority of contracts that programmers write in practice are short and simple. 
This means that, to be practical, methods and tools should make the best usage of such simple contracts or acquire more complex and complete specifications by other means (e.g., inference).
It also encourages the usage of simple specifications early on in the curriculum and in the training of programmers~\cite{KiniryZ08}.
\item
In spite of the simplicity of the contracts that are used in practice, developers who commit to using contracts seem to stick to them over an entire project lifetime.
This reveals that even simple specifications bring a value that is worth the effort: a little specification can go a long way.
\item
Developers often seem to adapt their contracts in response to changes in the design; future work in the direction of facilitating these adaptations and making them seamless has a potential for a high impact.
\item
A cornerstone software engineering principle---stable interfaces over changing implementations---seems to have been incorporated by programmers.
An interesting follow-up question is then whether this principle can be leveraged to improve not only the reusability of software components but also the \emph{collaboration} between programmers in a development team.
\item
Somewhat surprisingly, inheritance does not seem to affect most qualitative findings of our study.
The related important issue of how \emph{behavioral subtyping} is achieved in practice~\cite{PradelG13} belongs to future work, together with several other follow-up questions whose answers can build upon the foundations laid by this paper's results.
\end{inparaenum}

\extendedOnly{
\section{Conclusions} \label{sec:conclusion}
This paper presented an extensive empirical study of the evolution of specifications in the form of contracts.
The study targeted 15 projects written in Eiffel and C\# (using Code Contracts) over many years of development.
The main results show that the percentages of routines with pre- or postcondition and of classes with invariants is above 33\% for most projects; 
that these percentages tend to be stable over time, if we discount special events like merge or major refactorings; and that specifications change much less often than implementations---which makes a good case for stable interfaces over changing implementations.
}

\vskip2pt
\fakepar{Acknowledgments.} Thanks to Sebastian Nanz for comments on a draft of this paper; and to Todd Schiller, Kellen Donohue, Forrest Coward, and Mike Ernst for sharing a draft of their paper~\cite{writingcontracts} and comments on this work.

\bibliographystyle{splncs03}

\input{main.bbl}
\begin{sidewaystable}

\begin{minipage}{0.98\textwidth}
\begin{center}
\rowcolors{1}{}{lightgray}
{\scriptsize
\begin{tabular}{|l|rrrr|rrrr|rrrr|rrrr|rrrr|rrrr|rrrr|rrrr|}
  \hline
 &
\multicolumn{4}{c|}{\textsc{\# classes}} &
\multicolumn{4}{c|}{\textsc{\% classes inv}} &
\multicolumn{4}{c|}{\textsc{\# routines}} &
\multicolumn{4}{c|}{\textsc{\% routines spec}} &
\multicolumn{4}{c|}{\textsc{\% routines pre}} &
\multicolumn{4}{c|}{\textsc{\% routines post}} &
\multicolumn{4}{c|}{\textsc{avg routines pre}} &
\multicolumn{4}{c|}{\textsc{avg routines post}} 
\\
\textbf{Project} & \mm & \mmedian & \MM & \std & \mm & \mmedian & \MM & \std & \mm & \mmedian & \MM & \std & \mm & \mmedian & \MM & \std & \mm & \mmedian & \MM & \std & \mm & \mmedian & \MM & \std & \mm & \mmedian & \MM & \std & \mm & \mmedian & \MM & \std  \\ 
  \hline
AutoTest & 98 & 220 & 254 & 66 & 0.38 & 0.43 & 0.55 & 0.06 & 352 & 1053 & 1234 & 372 & 0.47 & 0.49 & 0.61 & 0.06 & 0.23 & 0.25 & 0.4 & 0.07 & 0.34 & 0.36 & 0.45 & 0.04 & 1.73 & 1.76 & 1.85 & 0.03 & 1.19 & 1.22 & 1.28 & 0.03  \\ 
  EiffelBase & 93 & 184 & 256 & 36 & 0.24 & 0.34 & 0.39 & 0.03 & 545 & 1984 & 3323 & 696 & 0.26 & 0.4 & 0.44 & 0.04 & 0.17 & 0.27 & 0.3 & 0.03 & 0.14 & 0.24 & 0.26 & 0.03 & 1.43 & 1.6 & 1.7 & 0.05 & 1.2 & 1.46 & 1.51 & 0.06  \\ 
  EiffelProgramAnalysis & 0 & 179 & 221 & 30 & 0 & 0.04 & 0.05 & 0 & 0 & 828 & 1127 & 199 & 0 & 0.25 & 0.27 & 0.02 & 0 & 0.14 & 0.16 & 0.02 & 0 & 0.15 & 0.16 & 0.01 & 0 & 1.23 & 1.25 & 0.09 & 0 & 1.13 & 1.17 & 0.08  \\ 
  GoboKernel & 0 & 72 & 157 & 38 & 0 & 0.11 & 0.13 & 0.04 & 0 & 168 & 702 & 155 & 0 & 0.6 & 1 & 0.17 & 0 & 0.3 & 0.4 & 0.09 & 0 & 0.51 & 1 & 0.19 & 0 & 2.1 & 2.91 & 0.59 & 0 & 1.32 & 1.86 & 0.25  \\ 
  GoboStructure & 42 & 75 & 109 & 17 & 0.19 & 0.33 & 0.39 & 0.06 & 122 & 372 & 483 & 88 & 0.18 & 0.29 & 0.41 & 0.07 & 0.07 & 0.19 & 0.28 & 0.06 & 0.16 & 0.23 & 0.32 & 0.05 & 1.45 & 1.82 & 1.93 & 0.13 & 1.17 & 1.44 & 1.49 & 0.1  \\ 
  GoboTime & 0 & 22 & 47 & 10 & 0 & 0.12 & 0.28 & 0.09 & 0 & 176 & 333 & 53 & 0 & 0.63 & 0.66 & 0.06 & 0 & 0.28 & 0.33 & 0.03 & 0 & 0.58 & 0.6 & 0.06 & 0 & 1.62 & 1.7 & 0.15 & 0 & 2.28 & 2.53 & 0.25  \\ 
  GoboUtility & 3 & 25 & 43 & 10 & 0 & 0.22 & 0.5 & 0.08 & 1 & 90 & 185 & 55 & 0 & 0.9 & 0.98 & 0.14 & 0 & 0.58 & 0.83 & 0.12 & 0 & 0.58 & 0.67 & 0.11 & 0 & 1.8 & 2.07 & 0.24 & 0 & 1.29 & 1.52 & 0.25  \\ 
  GoboXML & 0 & 176 & 859 & 252 & 0 & 0.38 & 0.48 & 0.07 & 0 & 883 & 5465 & 1603 & 0 & 0.35 & 0.44 & 0.05 & 0 & 0.23 & 0.35 & 0.03 & 0 & 0.23 & 0.33 & 0.06 & 0 & 1.43 & 1.55 & 0.14 & 0 & 1.2 & 1.36 & 0.07  \\ 
  Boogie & 9 & 606 & 647 & 181 & 0.24 & 0.34 & 0.58 & 0.06 & 80 & 3542 & 3748 & 1055 & 0.49 & 0.52 & 0.81 & 0.09 & 0.28 & 0.3 & 0.74 & 0.13 & 0.08 & 0.32 & 0.38 & 0.04 & 1.6 & 1.73 & 1.76 & 0.03 & 1 & 1.02 & 1.02 & 0.01  \\ 
  CCI & 45 & 60 & 108 & 15 & 0.01 & 0.04 & 0.06 & 0.01 & 160 & 210 & 302 & 50 & 0 & 0.03 & 0.05 & 0.01 & 0 & 0.03 & 0.04 & 0.01 & 0 & 0 & 0.01 & 0 & 1 & 1.33 & 1.6 & 0.22 & 0 & 0 & 1 & 0.49  \\ 
  Dafny & 11 & 148 & 184 & 25 & 0.04 & 0.47 & 0.52 & 0.06 & 25 & 375 & 551 & 85 & 0.16 & 0.64 & 0.74 & 0.07 & 0.16 & 0.57 & 0.64 & 0.06 & 0 & 0.18 & 0.22 & 0.03 & 1 & 2.29 & 2.36 & 0.18 & 0 & 1.04 & 1.05 & 0.14  \\ 
  Labs & 47 & 58 & 75 & 8 & 0.35 & 0.38 & 0.42 & 0.02 & 351 & 413 & 518 & 29 & 0.38 & 0.47 & 0.5 & 0.03 & 0.28 & 0.38 & 0.42 & 0.03 & 0.1 & 0.13 & 0.21 & 0.03 & 1.34 & 1.37 & 1.58 & 0.08 & 1.13 & 1.17 & 1.28 & 0.05  \\ 
  Quickgraph & 228 & 260 & 336 & 27 & 0 & 0.02 & 0.04 & 0.01 & 1074 & 1262 & 1862 & 179 & 0 & 0.16 & 0.22 & 0.07 & 0 & 0.15 & 0.21 & 0.07 & 0 & 0.01 & 0.02 & 0.01 & 0 & 1.71 & 2.1 & 0.71 & 0 & 1.18 & 1.36 & 0.46  \\ 
  Rxx & 0 & 145 & 189 & 53 & 0 & 0.42 & 0.44 & 0.08 & 0 & 1358 & 1792 & 494 & 0 & 0.7 & 0.97 & 0.11 & 0 & 0.6 & 0.93 & 0.13 & 0 & 0.62 & 0.81 & 0.08 & 0 & 2.1 & 2.24 & 0.18 & 0 & 1.03 & 1.12 & 0.1 \\ 
  Shweet & 0 & 28 & 36 & 13 & 0 & 0 & 0 & 0 & 0 & 57 & 85 & 33 & 0 & 0.1 & 0.4 & 0.07 & 0 & 0.1 & 0.4 & 0.07 & 0 & 0.01 & 0.07 & 0.02 & 0 & 1.6 & 2 & 0.77 & 0 & 1 & 1 & 0.49 \\
  DirectVCGen & 13 & 55 & 82 & 17 & 0 & 0 & 0.03 & 0 & 74 & 440 & 582 & 115 & 0.06 & 0.15 & 0.37 & 0.04 & 0.06 & 0.15 & 0.37 & 0.04 & 0.02 & 0.1 & 0.35 & 0.05 & 1 & 1 & 1.33 & 0.05 & 1 & 1 & 1 & 0\\ 
  ESCJava & 66 & 161 & 308 & 80 & 0.11 & 0.17 & 0.26 & 0.05 & 233 & 585 & 3079 & 853 & 0.16 & 0.36 & 0.74 & 0.21 & 0.14 & 0.27 & 0.69 & 0.2 & 0.06 & 0.12 & 0.2 & 0.03 & 1.07 & 1.27 & 1.66 & 0.21 & 1.21 & 1.52 & 1.88 & 0.12\\ 
  JavaFE & 107 & 124 & 641 & 29 & 0.12 & 0.47 & 0.62 & 0.04 & 499 & 589 & 1081 & 125 & 0.34 & 0.43 & 0.8 & 0.15 & 0.26 & 0.34 & 0.74 & 0.14 & 0.13 & 0.18 & 0.31 & 0.04 & 1.2 & 1.54 & 1.61 & 0.12 & 1.26 & 1.48 & 1.82 & 0.09\\ 
  Logging & 20 & 22 & 23 & 1 & 0.04 & 0.09 & 0.09 & 0.01 & 154 & 171 & 173 & 6 & 0.32 & 0.49 & 0.54 & 0.04 & 0.14 & 0.33 & 0.35 & 0.04 & 0.21 & 0.28 & 0.33 & 0.02 & 1.39 & 1.43 & 1.5 & 0.04 & 1.58 & 1.75 & 2 & 0.08\\ 
  RCC & 48 & 142 & 144 & 42 & 0.08 & 0.1 & 0.11 & 0.01 & 359 & 441 & 447 & 35 & 0.06 & 0.56 & 0.59 & 0.24 & 0.03 & 0.07 & 0.1 & 0.02 & 0.04 & 0.52 & 0.54 & 0.23 & 1.21 & 1.28 & 1.36 & 0.04 & 1 & 1.04 & 1.05 & 0.02\\ 
  Umbra & 23 & 41 & 77 & 16 & 0 & 0.06 & 0.1 & 0.03 & 36 & 122 & 332 & 78 & 0 & 0.02 & 0.05 & 0.02 & 0 & 0.01 & 0.03 & 0.01 & 0 & 0.02 & 0.04 & 0.01 & 0 & 1 & 1 & 0.49 & 0 & 1 & 1 & 0.47\\  
   \hline
\end{tabular}
}
\caption{Specification overall statistics with non-flat classes. For each project, we report the number of classes and of public routines (\textsc{\#~classes}, \textsc{\#~routines}); the percentage (1 is 100\%) of classes with non-empty invariant (\textsc{\%~classes inv}); of routines with non-empty specification (\textsc{\% routines spec}) and more specifically with non-empty precondition (\textsc{pre}) and postcondition (\textsc{post}); the mean number of clauses of routine preconditions (\textsc{avg routines pre}) and of postconditions (\textsc{post}). For each measure, the table reports minimum ($\mm$), median ($\mmedian$), maximum ($\MM$), and standard deviation ($\std$) across all revisions.}
\label{tab:overalls-inpaper}
\end{center}
\end{minipage} 

\begin{minipage}{0.98\textwidth}
\begin{tabular}{cccc}
	\includegraphics[width=4.8cm,height=3.7cm]{\figDir/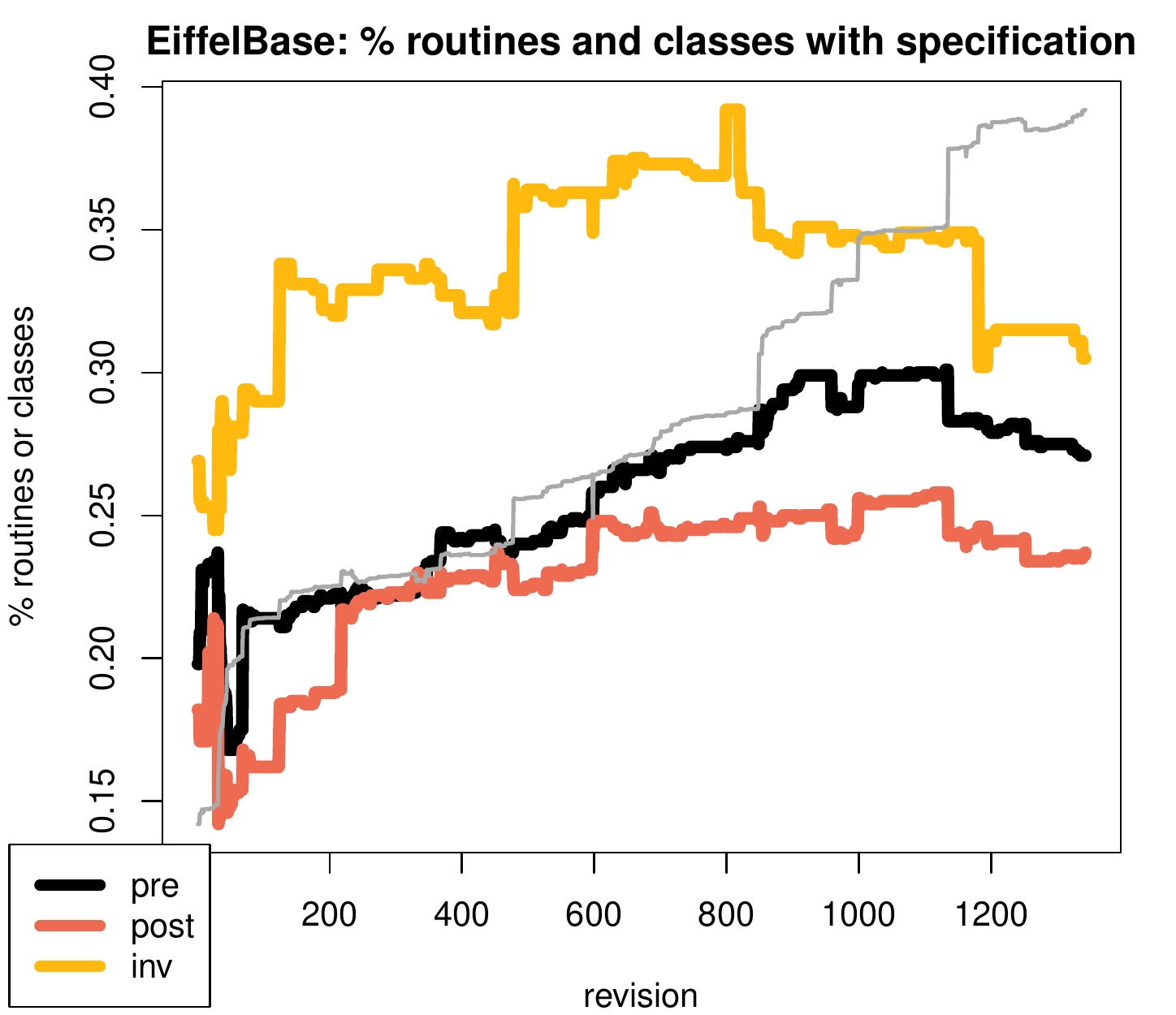} &
	\includegraphics[width=4.8cm,height=3.7cm]{\figDir/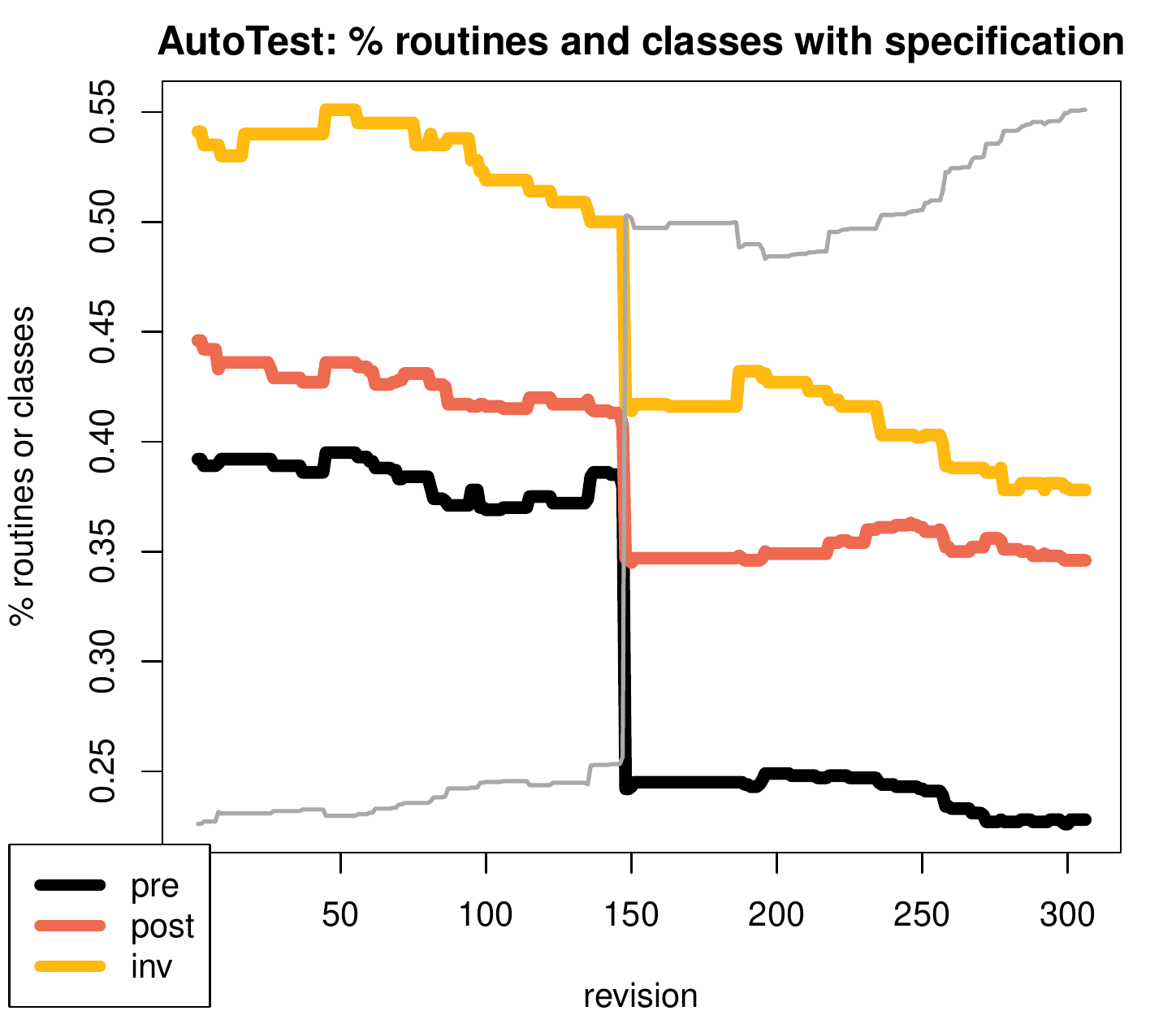} &
	\includegraphics[width=4.8cm,height=3.7cm]{\figDir/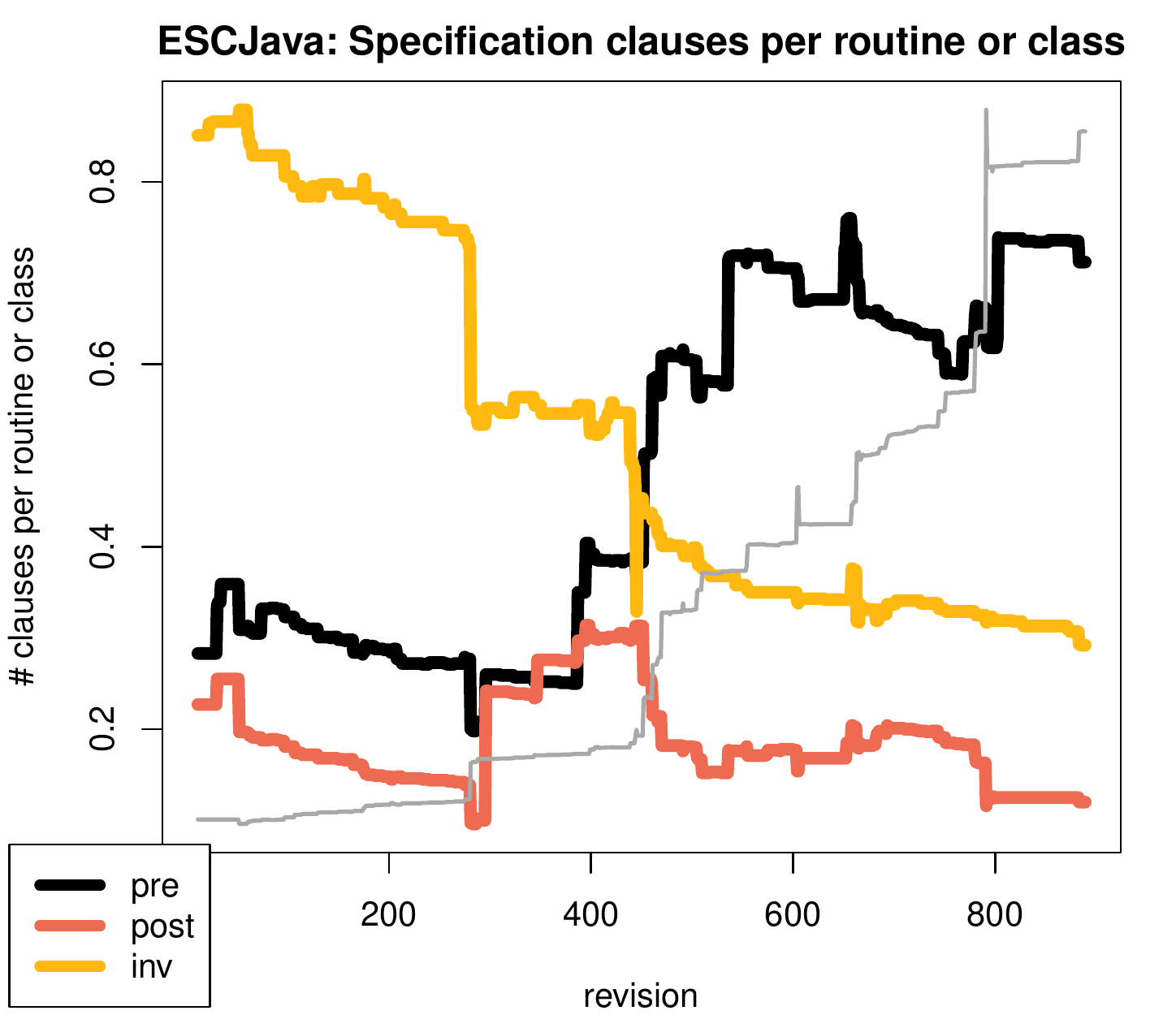} &
	\includegraphics[width=4.8cm,height=3.7cm]{\figDir/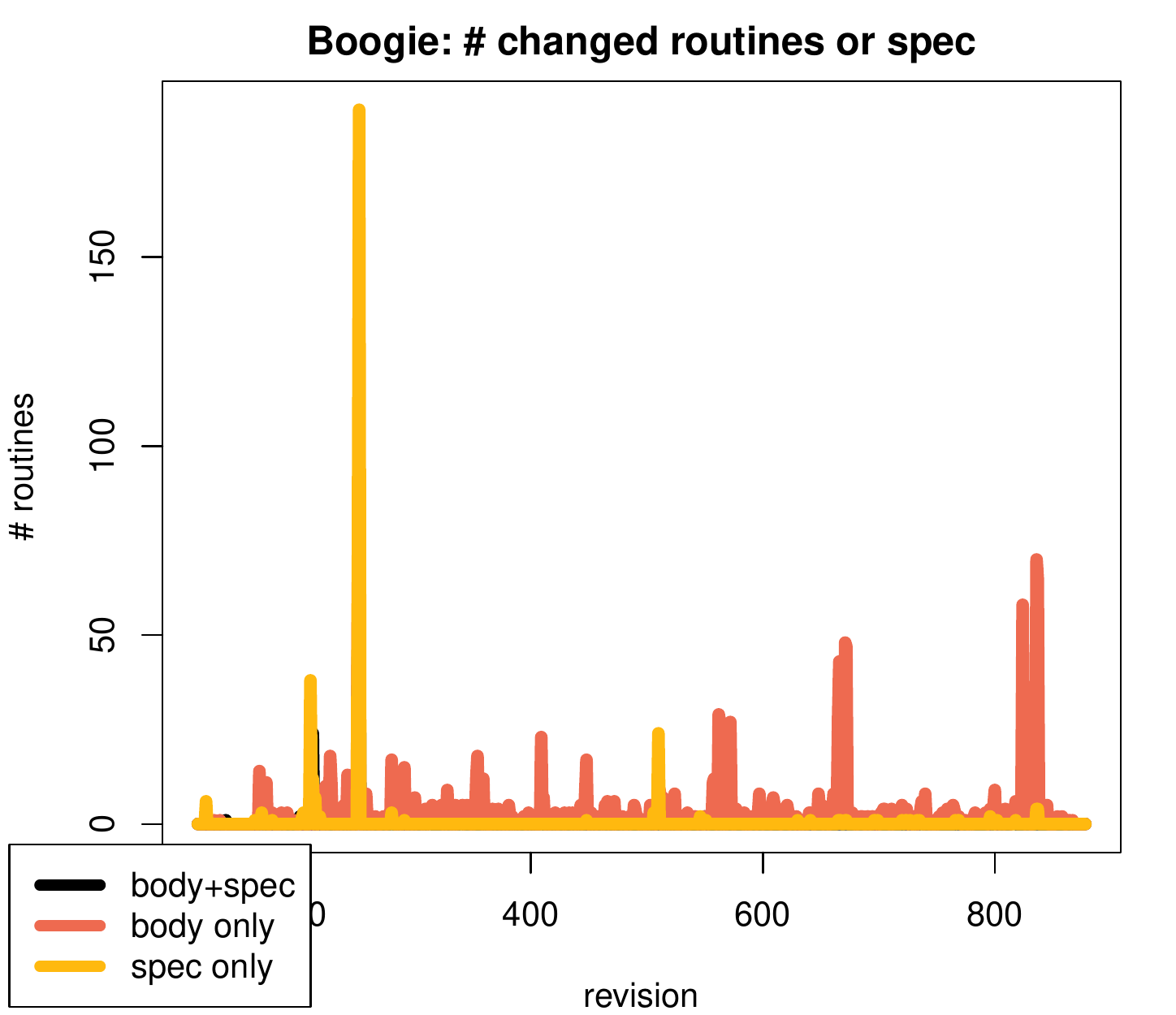} \\
\end{tabular}
\caption{
Selected plots for projects EiffelBase, AutoTest, ESCJava, and Boogie. Each graph from left to right represents the evolution over successive revisions of: (1) and (2), percentage of routines with precondition (\emph{pre} in the legend), with postcondition (\emph{post}), and of classes with invariant (\emph{inv}); (3), average number of clauses in contracts; (4), number of changes to implementation and specification (\emph{body$+$spec}), to implementation only (\emph{body only}), and change to \emph{spec}ification \emph{only}. 
When present, a thin gray line plots the total number of routine in the project (scaled). 
Similar plots for all projects are available~\cite{HowSpecChange-TR-17082012,COAT-data}.
}
\label{tab:eiffelbase-plots}
\end{minipage}

\end{sidewaystable}

\extendedOnly{
\clearpage
\newpage
\section{Appendix: Complete Statistics}
This appendix contains several tables with all statistics discussed in the paper.
In all tables, the few missing data about project Shweet are due to the fact that the project lacks class invariants, and hence the corresponding statistics are immaterial.

\fakepar{General specification statistics.}
Table~\ref{tab:overalls-avg} lists various general statistics about specifications: \# of classes, \% of classes with invariant, \# of routines, \% of routines with specification (pre- or postcondition), \% of routines with precondition, \% of routines with postcondition, average number of clauses in preconditions, average number of clauses in postconditions, average number of clauses in class invariants.
Table~\ref{tab:overalls-avg-flat} lists the same data but for flat classes.

Table~\ref{tab:overall-nonflat} lists other statistics about projects: language, number of revisions, age in weeks, lines of code, and then some of the same statistics about classes and routines as in Table~\ref{tab:overalls-avg}.
Table~\ref{tab:overall-flat} lists the same data for flat classes.

\fakepar{Change correlation analysis.}
Table~\ref{tab:change-major} lists Wilcoxon signed-rank tests about \linebreak changes, as described in Section~\ref{sec:impl-vs-spec}, comparing: changing and non-changing specifications of routines whose body  changes; changing and non-changing NE (i.e., non-empty) specifications of routines whose body chang\-es; preconditions becoming weaker vs.\ becoming stronger, NE preconditions becoming weaker vs.\ becoming stronger, postconditions becoming weaker vs.\ becoming stronger, NE postconditions becoming weaker vs.\ becoming stronger, class invariants becoming weaker vs.\ becoming stronger, NE class invariants becoming weaker vs.\ becoming stronger, class invariants becoming weaker when attributes are added vs.\ it not changing in strength; class invariants becoming stronger when attributes are added vs.\ it not changing in strength; class invariants becoming weaker when attributes are removed vs.\ it not changing in strength; class invariants becoming stronger when attributes are removed vs.\ it not changing in strength.
The statistics are $V$ and $p$ from the signed-rank test; $\Delta(\mu)$ is the difference in medians, whose value is positive iff the first---between the two compared measures---median is larger; $d$ is Cohen's effect size ($(m_1 - m_2)/\sigma$ where $m_1$, $m_2$ are the means of the two compared measures and $\sigma$ is the standard deviation of the whole measured data), whose value is positive iff the first---between the two compared measures---mean is larger.
The top half of the table considers non-flat classes, whereas the bottom half considers flat classes.
Table~\ref{tab:change-sum} shows the results of the same analysis but done by summing all changes instead of counting them with a binary value for each revision (see Section~\ref{sec:impl-vs-spec}).

\fakepar{Flat vs.\ non-flat correlation analysis.}
Table~\ref{tab:correlate-flatnonflat} lists correlation statistics between the evolution of the following measures in flat and non-flat classes: \% of routines with specification; \% of classes with invariants; average number of precondition clauses; average number of postcondition clauses, average number of invariant clauses.

\fakepar{Size correlation analysis.}
Table~\ref{tab:correlate-spec} lists correlation statistics between several pair of measures: \% of routines with specification and total \# of routines; \# of routines with specification and total \# of routines; \% of classes with invariant and total \# of classes; \# of classes with invariant and total \# of classes, average number of precondition clauses and \% of routines with precondition; average number of postcondition clauses and \% of routines with postcondition; average number of invariant clauses and \% of classes with invariants.
Table~\ref{tab:correlate-spec-flat} lists the same statistics for flat classes.

\fakepar{Public vs.\ non-public correlation analysis.}
Table~\ref{tab:correlate-privatepublic} lists correlation statistics between: \% of public routines with precondition and \% of non-public routines with precondition; \% of public routines with postcondition and \% of non-public routines with postcondition; \% of public routines with specification and \% of non-public routines with specification.
Table~\ref{tab:correlate-privatepublic-flat} lists the same statistics for flat classes.

\fakepar{Kinds of constructs.}
Table~\ref{tab:contract-types} lists statistics about constructs used in contracts: \% of preconditions with \lstinline|Void| or \textbf{null} checks; \% of preconditions with some form of quantification; \% of postconditions with \lstinline|Void| or \textbf{null} checks; \% of postconditions with some form of quantification; \% of postconditions with \lstinline|old|; \% of class invariants with \lstinline|Void| or \textbf{null} checks; \% of class invariants with some form of quantification.

\clearpage
\newpage
\changepage{}{}{-5mm}{-5mm}{}{-10mm}{}{}{10mm}
\input{\tablesLoc/overalls-avg.tex}
\input{\tablesLoc/overalls-avg-flat.tex}

\clearpage
\newpage
\input{\tablesLoc/overall-nonflat.tex}
\input{\tablesLoc/overall-flat.tex}

\clearpage
\newpage
\input{\tablesLoc/change-major.tex}
\input{\tablesLoc/change-sum.tex}


\clearpage
\newpage
\input{\tablesLoc/correlate-flatnonflat.tex}

\clearpage
\newpage
\input{\tablesLoc/correlate-spec.tex}
\input{\tablesLoc/correlate-spec-flat.tex}

\clearpage
\newpage
\input{\tablesLoc/correlate-privatepublic.tex}
\input{\tablesLoc/correlate-privatepublic-flat.tex}

\clearpage
\newpage
\input{\tablesLoc/contract-types.tex}

\clearpage
\newpage
\input{graphReport.tex}

\clearpage
\newpage
\input{dose-data.tex}

}

\end{document}

%% file: overalls-avg.tex
\begin{sidewaystable}[ht]
\begin{center}
{\tiny
\begin{tabular}{|l|rrrr|rrrr|rrrr|rrrr|rrrr|rrrr|rrrr|rrrr|rrrr|}
  \hline
 &
\multicolumn{4}{c|}{\textsc{\# classes}} &
\multicolumn{4}{c|}{\textsc{\% classes inv}} &
\multicolumn{4}{c|}{\textsc{\# routines}} &
\multicolumn{4}{c|}{\textsc{\% routines spec}} &
\multicolumn{4}{c|}{\textsc{\% routines pre}} &
\multicolumn{4}{c|}{\textsc{\% routines post}} &
\multicolumn{4}{c|}{\textsc{avg routines pre}} &
\multicolumn{4}{c|}{\textsc{avg routines post}} &
\multicolumn{4}{c|}{\textsc{avg inv}} 
\\
\textbf{Project} & \mm & \mmedian & \MM & \std & \mm & \mmedian & \MM & \std & \mm & \mmedian & \MM & \std & \mm & \mmedian & \MM & \std & \mm & \mmedian & \MM & \std & \mm & \mmedian & \MM & \std & \mm & \mmedian & \MM & \std & \mm & \mmedian & \MM & \std & \mm & \mmedian & \MM & \std \\ 
  \hline
AutoTest & 98 & 220 & 254 & 66 & 0.38 & 0.43 & 0.55 & 0.06 & 352 & 1053 & 1234 & 372 & 0.47 & 0.49 & 0.61 & 0.06 & 0.23 & 0.25 & 0.4 & 0.07 & 0.34 & 0.36 & 0.45 & 0.04 & 1.73 & 1.76 & 1.85 & 0.03 & 1.19 & 1.22 & 1.28 & 0.03 & 2.76 & 3.07 & 3.51 & 0.22 \\ 
  EiffelBase & 93 & 184 & 256 & 36 & 0.24 & 0.34 & 0.39 & 0.03 & 545 & 1984 & 3323 & 696 & 0.26 & 0.4 & 0.44 & 0.04 & 0.17 & 0.27 & 0.3 & 0.03 & 0.14 & 0.24 & 0.26 & 0.03 & 1.43 & 1.6 & 1.7 & 0.05 & 1.2 & 1.46 & 1.51 & 0.06 & 2.08 & 2.3 & 2.62 & 0.13 \\ 
  EiffelProgramAnalysis & 0 & 179 & 221 & 30 & 0 & 0.04 & 0.05 & 0 & 0 & 828 & 1127 & 199 & 0 & 0.25 & 0.27 & 0.02 & 0 & 0.14 & 0.16 & 0.02 & 0 & 0.15 & 0.16 & 0.01 & 0 & 1.23 & 1.25 & 0.09 & 0 & 1.13 & 1.17 & 0.08 & 0 & 1.17 & 1.33 & 0.12 \\ 
  GoboKernel & 0 & 72 & 157 & 38 & 0 & 0.11 & 0.13 & 0.04 & 0 & 168 & 702 & 155 & 0 & 0.6 & 1 & 0.17 & 0 & 0.3 & 0.4 & 0.09 & 0 & 0.51 & 1 & 0.19 & 0 & 2.1 & 2.91 & 0.59 & 0 & 1.32 & 1.86 & 0.25 & 0 & 1.44 & 2 & 0.41 \\
\hline
\end{tabular}
}
\caption{Specification overall statistics with non-flattened classes.}
\label{tab:overalls-avg}
\end{center}
\end{sidewaystable}

%% file: overalls-avg-flat.tex
\begin{sidewaystable}[ht]
\begin{center}
{\tiny
\begin{tabular}{|l|rrrr|rrrr|rrrr|rrrr|rrrr|rrrr|rrrr|rrrr|rrrr|}
  \hline
 &
\multicolumn{4}{c|}{\textsc{\# classes}} &
\multicolumn{4}{c|}{\textsc{\% classes inv}} &
\multicolumn{4}{c|}{\textsc{\# routines}} &
\multicolumn{4}{c|}{\textsc{\% routines spec}} &
\multicolumn{4}{c|}{\textsc{\% routines pre}} &
\multicolumn{4}{c|}{\textsc{\% routines post}} &
\multicolumn{4}{c|}{\textsc{avg routines pre}} &
\multicolumn{4}{c|}{\textsc{avg routines post}} &
\multicolumn{4}{c|}{\textsc{avg inv}} 
\\
\textbf{Project} & \mm & \mmedian & \MM & \std & \mm & \mmedian & \MM & \std & \mm & \mmedian & \MM & \std & \mm & \mmedian & \MM & \std & \mm & \mmedian & \MM & \std & \mm & \mmedian & \MM & \std & \mm & \mmedian & \MM & \std & \mm & \mmedian & \MM & \std & \mm & \mmedian & \MM & \std \\ 
  \hline
AutoTest & 98 & 220 & 254 & 66 & 0.48 & 0.54 & 0.63 & 0.05 & 524 & 2395 & 2946 & 1005 & 0.58 & 0.6 & 0.77 & 0.06 & 0.37 & 0.39 & 0.48 & 0.04 & 0.38 & 0.41 & 0.59 & 0.06 & 1.7 & 1.74 & 1.82 & 0.03 & 1.11 & 1.13 & 1.21 & 0.04 & 3.42 & 3.58 & 3.81 & 0.14 \\ 
  EiffelBase & 93 & 184 & 256 & 36 & 0.49 & 0.66 & 0.69 & 0.04 & 1377 & 5214 & 7542 & 1326 & 0.54 & 0.68 & 0.69 & 0.03 & 0.38 & 0.48 & 0.49 & 0.02 & 0.3 & 0.45 & 0.49 & 0.03 & 1.3 & 1.38 & 1.45 & 0.03 & 1.27 & 1.49 & 1.53 & 0.04 & 6.75 & 10.8 & 13.27 & 1.16 \\ 
  EiffelProgramAnalysis & 0 & 179 & 221 & 30 & 0 & 0.07 & 0.08 & 0.01 & 0 & 5300 & 8360 & 2064 & 0 & 0.2 & 0.25 & 0.03 & 0 & 0.11 & 0.18 & 0.03 & 0 & 0.1 & 0.13 & 0.02 & 0 & 1.16 & 1.34 & 0.11 & 0 & 1.13 & 1.59 & 0.17 & 0 & 1.1 & 1.21 & 0.1 \\ 
  GoboKernel & 0 & 72 & 157 & 38 & 0 & 0.12 & 0.29 & 0.04 & 0 & 271 & 1304 & 308 & 0 & 0.84 & 1 & 0.15 & 0 & 0.49 & 0.6 & 0.16 & 0 & 0.62 & 1 & 0.18 & 0 & 2.26 & 3.31 & 0.65 & 0 & 1.43 & 1.66 & 0.24 & 0 & 1.4 & 2 & 0.41 \\
\hline
\end{tabular}
}
\caption{Specification overall statistics with flattened classes.}
\label{tab:overalls-avg-flat}
\end{center}
\end{sidewaystable}

%% file: overall-nonflat.tex
\begin{sidewaystable}[ht]
\begin{center}
{\scriptsize
\begin{tabular}{|l|lrr|rrrr|rrrr|rrrr|rrrr|rrrr|rrrr|rrrr|}
  \hline
 &
 &  &  &
\multicolumn{4}{c|}{\textsc{LOC}} &
\multicolumn{4}{c|}{\textsc{\# classes}} &
\multicolumn{4}{c|}{\textsc{\% classes inv}} &
\multicolumn{4}{c|}{\textsc{\# routines}} &
\multicolumn{4}{c|}{\textsc{\% routines spec}} &
\multicolumn{4}{c|}{\textsc{\% routines pre}} &
\multicolumn{4}{c|}{\textsc{\% routines post}} 
\\
\textbf{Project} & \textbf{lang} & \textbf{rev} & \textbf{age} & \mm & \mmedian & \MM & \std & \mm & \mmedian & \MM & \std & \mm & \mmedian & \MM & \std & \mm & \mmedian & \MM & \std & \mm & \mmedian & \MM & \std & \mm & \mmedian & \MM & \std & \mm & \mmedian & \MM & \std \\ 
  \hline
AutoTest & Eiffel & 306 & 195 & 17154 & 52958 & 65625 & 19291 & 98 & 220 & 254 & 66 & 0.38 & 0.43 & 0.55 & 0.06 & 352 & 1053 & 1234 & 372 & 0.47 & 0.49 & 0.61 & 0.06 & 0.23 & 0.25 & 0.4 & 0.07 & 0.34 & 0.36 & 0.45 & 0.04 \\ 
  EiffelBase & Eiffel & 1342 & 1006 & 13154 & 45200 & 61922 & 11922 & 93 & 184 & 256 & 36 & 0.24 & 0.34 & 0.39 & 0.03 & 545 & 1984 & 3323 & 696 & 0.26 & 0.4 & 0.44 & 0.04 & 0.17 & 0.27 & 0.3 & 0.03 & 0.14 & 0.24 & 0.26 & 0.03 \\ 
  EiffelProgramAnalysis & Eiffel & 208 & 114 & 0 & 27498 & 40750 & 8546 & 0 & 179 & 221 & 30 & 0 & 0.04 & 0.05 & 0 & 0 & 828 & 1127 & 199 & 0 & 0.25 & 0.27 & 0.02 & 0 & 0.14 & 0.16 & 0.02 & 0 & 0.15 & 0.16 & 0.01 \\ 
  GoboKernel & Eiffel & 671 & 747 & 0 & 8457 & 53316 & 9175 & 0 & 72 & 157 & 38 & 0 & 0.11 & 0.13 & 0.04 & 0 & 168 & 702 & 155 & 0 & 0.6 & 1 & 0.17 & 0 & 0.3 & 0.4 & 0.09 & 0 & 0.51 & 1 & 0.19 \\ 
  GoboStructure & Eiffel & 282 & 716 & 3995 & 13069 & 21941 & 4525 & 42 & 75 & 109 & 17 & 0.19 & 0.33 & 0.39 & 0.06 & 122 & 372 & 483 & 88 & 0.18 & 0.29 & 0.41 & 0.07 & 0.07 & 0.19 & 0.28 & 0.06 & 0.16 & 0.23 & 0.32 & 0.05 \\ 
  GoboTime & Eiffel & 120 & 524 & 0 & 4131 & 10840 & 1753 & 0 & 22 & 47 & 10 & 0 & 0.12 & 0.28 & 0.09 & 0 & 176 & 333 & 53 & 0 & 0.63 & 0.66 & 0.06 & 0 & 0.28 & 0.33 & 0.03 & 0 & 0.58 & 0.6 & 0.06 \\
\hline
\end{tabular}
}
\caption{Project specification statistics for non-flattened classes.}
\label{tab:overall-nonflat}
\end{center}
\end{sidewaystable}

%% file: overall-flat.tex
\begin{sidewaystable}[ht]
\begin{center}
{\scriptsize
\begin{tabular}{|l|lrr|rrrr|rrrr|rrrr|rrrr|rrrr|rrrr|rrrr|}
  \hline
 &
 &  &  &
\multicolumn{4}{c|}{\textsc{LOC}} &
\multicolumn{4}{c|}{\textsc{\# classes}} &
\multicolumn{4}{c|}{\textsc{\% classes inv}} &
\multicolumn{4}{c|}{\textsc{\# routines}} &
\multicolumn{4}{c|}{\textsc{\% routines spec}} &
\multicolumn{4}{c|}{\textsc{\% routines pre}} &
\multicolumn{4}{c|}{\textsc{\% routines post}} 
\\
\textbf{Project} & \textbf{lang} & \textbf{rev} & \textbf{age} & \mm & \mmedian & \MM & \std & \mm & \mmedian & \MM & \std & \mm & \mmedian & \MM & \std & \mm & \mmedian & \MM & \std & \mm & \mmedian & \MM & \std & \mm & \mmedian & \MM & \std & \mm & \mmedian & \MM & \std \\ 
  \hline
AutoTest & Eiffel & 306 & 195 & 17154 & 52958 & 65625 & 19291 & 98 & 220 & 254 & 66 & 0.48 & 0.54 & 0.63 & 0.05 & 524 & 2395 & 2946 & 1005 & 0.58 & 0.6 & 0.77 & 0.06 & 0.37 & 0.39 & 0.48 & 0.04 & 0.38 & 0.41 & 0.59 & 0.06 \\ 
  EiffelBase & Eiffel & 1342 & 1006 & 13154 & 45200 & 61922 & 11922 & 93 & 184 & 256 & 36 & 0.49 & 0.66 & 0.69 & 0.04 & 1377 & 5214 & 7542 & 1326 & 0.54 & 0.68 & 0.69 & 0.03 & 0.38 & 0.48 & 0.49 & 0.02 & 0.3 & 0.45 & 0.49 & 0.03 \\ 
  EiffelProgramAnalysis & Eiffel & 208 & 114 & 0 & 27498 & 40750 & 8546 & 0 & 179 & 221 & 30 & 0 & 0.07 & 0.08 & 0.01 & 0 & 5300 & 8360 & 2064 & 0 & 0.2 & 0.25 & 0.03 & 0 & 0.11 & 0.18 & 0.03 & 0 & 0.1 & 0.13 & 0.02 \\ 
  GoboKernel & Eiffel & 671 & 747 & 0 & 8457 & 53316 & 9175 & 0 & 72 & 157 & 38 & 0 & 0.12 & 0.29 & 0.04 & 0 & 271 & 1304 & 308 & 0 & 0.84 & 1 & 0.15 & 0 & 0.49 & 0.6 & 0.16 & 0 & 0.62 & 1 & 0.18 \\ 
  GoboStructure & Eiffel & 282 & 716 & 3995 & 13069 & 21941 & 4525 & 42 & 75 & 109 & 17 & 0.5 & 0.81 & 0.87 & 0.11 & 245 & 1421 & 2900 & 696 & 0.7 & 0.83 & 0.91 & 0.04 & 0.51 & 0.57 & 0.74 & 0.05 & 0.53 & 0.61 & 0.76 & 0.06 \\ 
  GoboTime & Eiffel & 120 & 524 & 0 & 4131 & 10840 & 1753 & 0 & 22 & 47 & 10 & 0 & 0.18 & 0.38 & 0.1 & 0 & 346 & 630 & 87 & 0 & 0.85 & 0.88 & 0.08 & 0 & 0.46 & 0.53 & 0.05 & 0 & 0.69 & 0.73 & 0.07 \\
\hline
\end{tabular}
}
\caption{Project specification statistics for flattened classes.}
\label{tab:overall-flat}
\end{center}
\end{sidewaystable}

%% file: change-major.tex
\begin{table*}[ht]
\begin{center}
{\scriptsize
\begin{tabular}{|c|rrrrrrrrrrrr|}
  \hline
&
\begin{sideways} chang vs unch spec \end{sideways} &
\begin{sideways} NE chang vs unch spec \end{sideways} &
\begin{sideways} pre weak vs strong \end{sideways} &
\begin{sideways} NE pre weak vs strong \end{sideways} &
\begin{sideways} post weak vs strong \end{sideways} &
\begin{sideways} NE post weak vs strong \end{sideways} &
\begin{sideways} inv weak vs strong \end{sideways} &
\begin{sideways} NE inv weak vs strong \end{sideways} &
\begin{sideways} $+$A inv weak vs same \end{sideways} &
\begin{sideways} $+$A inv strong vs same \end{sideways} &
\begin{sideways} $-$A inv weak vs same \end{sideways} &
\begin{sideways} $-$A inv strong vs same \end{sideways} \\

  \hline
$V$ & 0.00E+00 & 2.90E+01 & 3.15E+01 & 1.90E+01 & 2.90E+01 & 0.00E+00 & 6.15E+01 & 5.85E+01 & 0.00E+00 & 0.00E+00 & 3.00E+00 & 0.00E+00 \\ 
  $p$ & 9.54E-07 & 4.78E-03 & 1.95E-02 & 3.78E-02 & 1.44E-02 &  & 1.82E-01 & 6.40E-01 & 1.41E-02 & 1.40E-02 & 1.41E-01 & 2.23E-02 \\ 
  $\Delta(\mu)$ & -6.80E+01 & -1.10E+01 & -2.00E+00 & 0.00E+00 & -2.00E+00 & 0.00E+00 & -1.00E+00 & 0.00E+00 & 0.00E+00 & 0.00E+00 & 0.00E+00 & 0.00E+00 \\ 
  $d$ & -9.96E-01 & -5.03E-01 & -4.04E-01 & -4.75E-01 & -4.19E-01 & 0.00E+00 & -1.82E-01 & -1.29E-01 & -6.27E-01 & -4.17E-01 & -3.09E-01 & -5.25E-01 \\ 
   \hline
$V$ & 0.00E+00 & 2.90E+01 & 3.15E+01 & 1.90E+01 & 2.90E+01 & 0.00E+00 & 6.15E+01 & 5.85E+01 & 0.00E+00 & 0.00E+00 & 3.00E+00 & 0.00E+00 \\ 
  $p$ & 9.54E-07 & 4.78E-03 & 1.95E-02 & 3.78E-02 & 1.44E-02 &  & 1.82E-01 & 6.40E-01 & 1.41E-02 & 1.40E-02 & 1.41E-01 & 2.23E-02 \\ 
  $\Delta(\mu)$ & -6.80E+01 & -1.10E+01 & -2.00E+00 & 0.00E+00 & -2.00E+00 & 0.00E+00 & -1.00E+00 & 0.00E+00 & 0.00E+00 & 0.00E+00 & 0.00E+00 & 0.00E+00 \\ 
  $d$ & -9.96E-01 & -5.03E-01 & -4.04E-01 & -4.75E-01 & -4.19E-01 & 0.00E+00 & -1.82E-01 & -1.29E-01 & -6.27E-01 & -4.17E-01 & -3.09E-01 & -5.25E-01 \\ 
   \hline
\end{tabular}
}
\caption{Change analysis with majority for non-flattened (top) and flattened (bottom) classes.}
\label{tab:change-major}
\end{center}
\end{table*}

%% file: change-sum.tex
\begin{table*}[ht]
\begin{center}
{\scriptsize
\begin{tabular}{|c|rrrrrrrrrrrr|}
  \hline
&
\begin{sideways} chang vs unch spec \end{sideways} &
\begin{sideways} NE chang vs unch spec \end{sideways} &
\begin{sideways} pre weak vs strong \end{sideways} &
\begin{sideways} NE pre weak vs strong \end{sideways} &
\begin{sideways} post weak vs strong \end{sideways} &
\begin{sideways} NE post weak vs strong \end{sideways} &
\begin{sideways} inv weak vs strong \end{sideways} &
\begin{sideways} NE inv weak vs strong \end{sideways} &
\begin{sideways} $+$A inv weak vs same \end{sideways} &
\begin{sideways} $+$A inv strong vs same \end{sideways} &
\begin{sideways} $-$A inv weak vs same \end{sideways} &
\begin{sideways} $-$A inv strong vs same \end{sideways} \\

  \hline
$V$ & 0.00E+00 & 7.75E+01 & 5.45E+01 & 2.55E+01 & 5.00E+01 & 0.00E+00 & 4.35E+01 & 4.80E+01 & 0.00E+00 & 0.00E+00 & 3.50E+00 & 0.00E+00 \\ 
  $p$ & 9.54E-07 & 1.92E-01 & 1.07E-01 & 9.60E-02 & 4.15E-02 &  & 7.00E-02 & 5.12E-01 & 1.43E-02 & 1.41E-02 & 9.04E-02 & 2.23E-02 \\ 
  $\Delta(\mu)$ & -2.28E+02 & -8.00E+00 & -4.00E+00 & 0.00E+00 & -4.00E+00 & 0.00E+00 & -1.00E+00 & 0.00E+00 & 0.00E+00 & 0.00E+00 & 0.00E+00 & 0.00E+00 \\ 
  $d$ & -8.68E-01 & -2.61E-01 & -2.83E-01 & -5.85E-01 & -3.20E-01 & 0.00E+00 & -3.32E-01 & -1.21E-01 & -5.91E-01 & -4.11E-01 & -3.34E-01 & -5.37E-01 \\ 
   \hline
$V$ & 0.00E+00 & 7.75E+01 & 5.45E+01 & 2.55E+01 & 5.00E+01 & 0.00E+00 & 4.35E+01 & 4.80E+01 & 0.00E+00 & 0.00E+00 & 3.50E+00 & 0.00E+00 \\ 
  $p$ & 9.54E-07 & 1.92E-01 & 1.07E-01 & 9.60E-02 & 4.15E-02 &  & 7.00E-02 & 5.12E-01 & 1.43E-02 & 1.41E-02 & 9.04E-02 & 2.23E-02 \\ 
  $\Delta(\mu)$ & -2.28E+02 & -8.00E+00 & -4.00E+00 & 0.00E+00 & -4.00E+00 & 0.00E+00 & -1.00E+00 & 0.00E+00 & 0.00E+00 & 0.00E+00 & 0.00E+00 & 0.00E+00 \\ 
  $d$ & -8.68E-01 & -2.61E-01 & -2.83E-01 & -5.85E-01 & -3.20E-01 & 0.00E+00 & -3.32E-01 & -1.21E-01 & -5.91E-01 & -4.11E-01 & -3.34E-01 & -5.37E-01 \\ 
   \hline
\end{tabular}
}
\caption{Change analysis for non-flattened (top) and flattened (bottom) classes.}
\label{tab:change-sum}
\end{center}
\end{table*}

%% file: correlate-flatnonflat.tex
\begin{sidewaystable}[ht]
\begin{center}
{\footnotesize
\begin{tabular}{|l|rr|rr|rr|rr|rr|}
  \hline
 &
\multicolumn{2}{|c}{\textsc{\% routines w/ spec}} &
\multicolumn{2}{|c}{\textsc{\% classes w/ inv}} &
\multicolumn{2}{|c}{\textsc{avg pre strength}} &
\multicolumn{2}{|c}{\textsc{avg post strength}} &
\multicolumn{2}{|c|}{\textsc{avg inv strength}}
\\
\textbf{Project} & $p$ & $\tau$ & $p$ & $\tau$ & $p$ & $\tau$ & $p$ & $\tau$ & $p$ & $\tau$ \\ 
  \hline
AutoTest & 0.00E+00 & 0.65 & 0.00E+00 & 0.98 & 0.00E+00 & 0.60 & 0.00E+00 & 0.75 & 1.55E-04 & 0.16 \\ 
  EiffelBase & 0.00E+00 & 0.17 & 0.00E+00 & 0.27 & 0.00E+00 & 0.62 & 0.00E+00 & 0.38 & 0.00E+00 & 0.46 \\ 
  EiffelProgramAnalysis & 8.98E-01 & -0.01 & 0.00E+00 & 0.66 & 4.75E-12 & 0.34 & 1.24E-10 & 0.31 & 0.00E+00 & 0.95 \\ 
  GoboKernel & 0.00E+00 & 0.60 & 0.00E+00 & 0.86 & 0.00E+00 & 0.89 & 0.00E+00 & 0.36 & 0.00E+00 & 0.97 \\ 
  GoboStructure & 8.88E-01 & 0.01 & 1.44E-03 & 0.13 & 0.00E+00 & 0.38 & 0.00E+00 & 0.70 & 0.00E+00 & 0.45 \\ 
  GoboTime & 4.28E-01 & -0.05 & 0.00E+00 & 0.81 & 0.00E+00 & 0.58 & 0.00E+00 & 0.89 & 1.41E-06 & 0.36 \\
\hline
\end{tabular}
}
\caption{Correlation of measures in flat and non-flat classes.}
\label{tab:correlate-flatnonflat}
\end{center}
\end{sidewaystable}

%% file: correlate-spec.tex
\begin{table*}[ht]
\begin{center}
{\scriptsize
\begin{tabular}{|l|rr|rr|rr|rr|rr|rr|rr|}
  \hline
 &
\multicolumn{2}{|c}{\textsc{\% spec/\# routines}} &
\multicolumn{2}{|c}{\textsc{\# spec/\# routines}} &
\multicolumn{2}{|c}{\textsc{\% inv/\# classes}} &
\multicolumn{2}{|c}{\textsc{\# inv/\# classes}} &
\multicolumn{2}{|c}{\textsc{avg/\% pre}} &
\multicolumn{2}{|c}{\textsc{avg/\% post}} &
\multicolumn{2}{|c|}{\textsc{avg/\% inv}}
\\
\textbf{Project} & $p$ & $\tau$ & $p$ & $\tau$ & $p$ & $\tau$ & $p$ & $\tau$ & $p$ & $\tau$ & $p$ & $\tau$ & $p$ & $\tau$ \\ 
  \hline
AutoTest & 1.02E-94 & -0.82 & 0.00E+00 & 0.93 & 1.21E-117 & -0.93 & 0.00E+00 & 0.76 & 0.00E+00 & 0.65 & 0.00E+00 & 0.43 & 0.00E+00 & 0.78 \\ 
  EiffelBase & 0.00E+00 & 0.66 & 0.00E+00 & 0.98 & 1.13E-08 & 0.11 & 0.00E+00 & 0.95 & 0.00E+00 & 0.44 & 0.00E+00 & 0.26 & 0.00E+00 & 0.19 \\ 
  EiffelProgramAnalysis & 1.17E-10 & 0.31 & 0.00E+00 & 0.98 & 8.25E-04 & -0.17 & 0.00E+00 & 0.85 & 1.69E-01 & 0.07 & 4.94E-01 & 0.03 & 5.23E-01 & -0.04 \\ 
  GoboKernel & 9.98E-04 & -0.09 & 0.00E+00 & 0.97 & 0.00E+00 & 0.35 & 0.00E+00 & 0.92 & 2.71E-05 & -0.11 & 1.42E-01 & -0.04 & 6.90E-11 & 0.18 \\ 
  GoboStructure & 3.04E-05 & 0.17 & 0.00E+00 & 0.53 & 5.69E-07 & -0.21 & 0.00E+00 & 0.70 & 4.23E-04 & 0.15 & 0.00E+00 & 0.52 & 6.58E-01 & 0.02 \\ 
  GoboTime & 2.16E-04 & -0.24 & 0.00E+00 & 0.91 & 6.16E-13 & 0.51 & 0.00E+00 & 0.91 & 1.55E-05 & 0.30 & 7.87E-10 & 0.41 & 6.79E-02 & 0.13 \\
\hline
\end{tabular}
}
\caption{Correlation of measures of specification in non-flat classes.}
\label{tab:correlate-spec}
\end{center}
\end{table*}

%% file: correlate-spec-flat.tex
\begin{table*}[ht]
\begin{center}
{\scriptsize
\begin{tabular}{|l|rr|rr|rr|rr|rr|rr|rr|}
  \hline
 &
\multicolumn{2}{|c}{\textsc{\% spec/\# routines}} &
\multicolumn{2}{|c}{\textsc{\# spec/\# routines}} &
\multicolumn{2}{|c}{\textsc{\% inv/\# classes}} &
\multicolumn{2}{|c}{\textsc{\# inv/\# classes}} &
\multicolumn{2}{|c}{\textsc{avg/\% pre}} &
\multicolumn{2}{|c}{\textsc{avg/\% post}} &
\multicolumn{2}{|c|}{\textsc{avg/\% inv}}
\\
\textbf{Project} & $p$ & $\tau$ & $p$ & $\tau$ & $p$ & $\tau$ & $p$ & $\tau$ & $p$ & $\tau$ & $p$ & $\tau$ & $p$ & $\tau$ \\ 
  \hline
AutoTest & 1.28E-70 & -0.70 & 0.00E+00 & 0.92 & 3.09E-117 & -0.93 & 0.00E+00 & 0.88 & 0.00E+00 & 0.51 & 0.00E+00 & 0.38 & 1.42E-01 & -0.06 \\ 
  EiffelBase & 8.85E-06 & -0.08 & 0.00E+00 & 0.92 & 9.35E-35 & -0.23 & 0.00E+00 & 0.94 & 0.00E+00 & 0.33 & 7.64E-24 & -0.19 & 0.00E+00 & 0.16 \\ 
  EiffelProgramAnalysis & 1.51E-14 & 0.37 & 0.00E+00 & 0.98 & 8.13E-19 & -0.44 & 0.00E+00 & 0.89 & 0.00E+00 & 0.50 & 4.93E-07 & 0.25 & 6.44E-13 & -0.39 \\ 
  GoboKernel & 0.00E+00 & 0.29 & 0.00E+00 & 0.99 & 0.00E+00 & 0.47 & 0.00E+00 & 0.92 & 7.60E-05 & -0.10 & 3.25E-64 & -0.45 & 4.99E-11 & 0.18 \\ 
  GoboStructure & 6.58E-01 & 0.02 & 0.00E+00 & 0.98 & 3.63E-13 & 0.30 & 0.00E+00 & 0.72 & 0.00E+00 & 0.53 & 8.23E-43 & -0.56 & 5.37E-14 & 0.31 \\ 
  GoboTime & 2.41E-01 & 0.08 & 0.00E+00 & 0.98 & 2.22E-16 & 0.58 & 0.00E+00 & 0.88 & 4.86E-05 & 0.27 & 1.87E-02 & 0.16 & 0.00E+00 & 0.80 \\
\hline
\end{tabular}
}
\caption{Correlation of measures of specification in flat classes.}
\label{tab:correlate-spec-flat}
\end{center}
\end{table*}

%% file: correlate-privatepublic.tex
\begin{table*}[ht]
\begin{center}
{\footnotesize
\begin{tabular}{|l|rr|rr|rr|}
  \hline
 &
\multicolumn{2}{|c}{\textsc{\% pre public/private}} &
\multicolumn{2}{|c}{\textsc{\% post public/private}} &
\multicolumn{2}{|c|}{\textsc{\% spec public/private}}
\\
\textbf{Project} & $p$ & $\tau$ & $p$ & $\tau$ & $p$ & $\tau$ \\ 
  \hline
AutoTest & 0.00E+00 & 0.76 & 0.00E+00 & 0.51 & 0.00E+00 & 0.75 \\ 
  EiffelBase & 0.00E+00 & 0.73 & 0.00E+00 & 0.57 & 0.00E+00 & 0.64 \\ 
  EiffelProgramAnalysis & 2.16E-09 & 0.29 & 0.00E+00 & 0.49 & 4.18E-01 & 0.04 \\ 
  GoboKernel & 1.17E-05 & -0.12 & 8.31E-01 & -0.01 & 1.48E-39 & -0.36 \\ 
  GoboStructure & 9.01E-27 & -0.44 & 1.05E-36 & -0.52 & 3.27E-30 & -0.47 \\ 
  GoboTime & 1.31E-03 & -0.22 & 2.35E-07 & 0.34 & 7.48E-05 & -0.26 \\ 
  GoboUtility & 1.59E-02 & 0.12 & 0.00E+00 & 0.65 & 3.59E-02 & 0.10 \\ 
  GoboXML & 1.57E-07 & -0.12 & 2.08E-02 & 0.05 & 2.66E-26 & -0.24 \\ 
  Boogie & 0.00E+00 & 0.94 & 0.00E+00 & 0.73 & 0.00E+00 & 0.94 \\ 
  CCI & 0.00E+00 & 0.84 & 0.00E+00 & 0.80 & 0.00E+00 & 0.85 \\ 
  Dafny & 0.00E+00 & 0.36 & 0.00E+00 & 0.86 & 0.00E+00 & 0.66 \\ 
  Labs & 1.66E-09 & -0.63 & 4.42E-03 & 0.30 & 2.20E-06 & -0.50 \\ 
  Quickgraph & 1.98E-07 & 0.19 & 0.00E+00 & 0.80 & 1.45E-07 & 0.19 \\ 
  Rxx & 7.66E-05 & 0.23 & 5.46E-06 & 0.26 & 2.95E-03 & 0.17 \\ 
  shweet & 6.28E-09 & 0.59 & 4.78E-05 & 0.42 & 3.11E-04 & 0.37 \\ 
  DirectVCGen & 1.97E-02 & -0.08 &  &  & 1.86E-02 & -0.08 \\ 
  ESCJava & 0.00E+00 & 0.40 & 7.38E-12 & 0.16 & 0.00E+00 & 0.46 \\ 
  JavaFE & 0.00E+00 & 0.33 & 1.56E-11 & 0.24 & 0.00E+00 & 0.51 \\ 
  Logging & 2.01E-08 & 0.84 & 1.45E-01 & 0.21 & 1.55E-08 & 0.82 \\ 
  RCC & 5.83E-02 & 0.26 & 2.22E-06 & 0.67 & 7.73E-07 & 0.68 \\ 
  Umbra & 0.00E+00 & 0.59 & 0.00E+00 & 0.78 & 0.00E+00 & 0.81 \\ 
   \hline
\textbf{Overall} & 0.00E+00 & 0.38 & 0.00E+00 & 0.38 & 0.00E+00 & 0.39 \\ 
   \hline
\end{tabular}
}
\caption{Correlation of measures between public and private routines in non-flat classes.}
\label{tab:correlate-privatepublic}
\end{center}
\end{table*}

%% file: correlate-privatepublic-flat.tex
\begin{table*}[ht]
\begin{center}
{\footnotesize
\begin{tabular}{|l|rr|rr|rr|}
  \hline
 &
\multicolumn{2}{|c}{\textsc{\% pre public/private}} &
\multicolumn{2}{|c}{\textsc{\% post public/private}} &
\multicolumn{2}{|c|}{\textsc{\% spec public/private}}
\\
\textbf{Project} & $p$ & $\tau$ & $p$ & $\tau$ & $p$ & $\tau$ \\ 
  \hline
AutoTest & 0.00E+00 & 0.66 & 0.00E+00 & 0.61 & 0.00E+00 & 0.76 \\ 
  EiffelBase & 0.00E+00 & 0.41 & 5.13E-18 & -0.16 & 8.27E-01 & 0.00 \\ 
  EiffelProgramAnalysis & 1.00E-06 & 0.24 & 0.00E+00 & 0.60 & 2.25E-10 & 0.31 \\ 
  GoboKernel & 7.45E-02 & 0.05 & 1.07E-08 & -0.15 & 3.82E-07 & -0.14 \\ 
  GoboStructure & 3.17E-01 & -0.04 & 0.00E+00 & 0.54 & 8.73E-06 & -0.18 \\ 
  GoboTime & 1.85E-03 & -0.21 & 3.49E-01 & -0.06 & 2.13E-06 & 0.32 \\ 
  GoboUtility & 1.39E-01 & 0.07 & 2.84E-04 & 0.18 & 1.19E-02 & 0.12 \\ 
  GoboXML & 0.00E+00 & 0.37 & 2.68E-01 & 0.02 & 0.00E+00 & 0.19 \\ 
  Boogie & 0.00E+00 & 0.68 & 0.00E+00 & 0.27 & 0.00E+00 & 0.84 \\ 
  CCI & 0.00E+00 & 0.83 & 0.00E+00 & 0.80 & 0.00E+00 & 0.83 \\ 
  Dafny & 0.00E+00 & 0.40 & 0.00E+00 & 0.80 & 0.00E+00 & 0.66 \\ 
  Labs & 2.89E-11 & -0.69 & 2.37E-02 & 0.23 & 3.94E-12 & -0.72 \\ 
  Quickgraph & 4.07E-08 & 0.20 & 0.00E+00 & 0.81 & 4.19E-09 & 0.21 \\ 
  Rxx & 1.01E-04 & 0.22 & 0.00E+00 & 0.47 & 4.77E-04 & 0.20 \\ 
  shweet & 9.05E-09 & 0.59 & 4.59E-05 & 0.43 & 7.61E-04 & 0.34 \\ 
  DirectVCGen & 1.10E-44 & -0.50 &  &  & 4.53E-43 & -0.49 \\ 
  ESCJava & 0.00E+00 & 0.50 & 5.61E-06 & 0.11 & 0.00E+00 & 0.55 \\ 
  JavaFE & 0.00E+00 & 0.60 & 5.17E-04 & 0.12 & 0.00E+00 & 0.59 \\ 
  Logging & 2.01E-08 & 0.84 & 1.45E-01 & 0.21 & 3.78E-08 & 0.80 \\ 
  RCC & 1.12E-01 & 0.22 & 6.58E-07 & 0.70 & 3.65E-07 & 0.70 \\ 
  Umbra & 0.00E+00 & 0.87 & 0.00E+00 & 0.77 & 0.00E+00 & 0.90 \\ 
   \hline
\textbf{Overall} & 0.00E+00 & 0.35 & 0.00E+00 & 0.50 & 0.00E+00 & 0.42 \\ 
   \hline
\end{tabular}
}
\caption{Correlation of measures between public and private routines in flat classes.}
\label{tab:correlate-privatepublic-flat}
\end{center}
\end{table*}

%% file: contract-types.tex
\begin{sidewaystable}[ht]
\begin{center}
{\scriptsize
%
}
\caption{Syntactic elements found in contracts (non-flattened classes).}
\label{tab:contract-types}
\end{center}
\end{sidewaystable}

%% file: graphReport.tex

\newcommand{\importPic}[2]{\includegraphics[width=5.4cm]{\allGraphsDir/#1/#1--#2.pdf}}
\newcommand{\importFlat}[2]{\includegraphics[width=5.4cm]{\allGraphsDir/#1/#1-#2.pdf}}

\newcommand{\printNonFlat}[2]{
\begin{figure}[!htb]
\begin{center}
    \begin{tabular}{cccc}
	\includegraphics[width=5.4cm]{\allGraphsDir/AutoTest_72/AutoTest_72--#1.pdf} &
	\includegraphics[width=5.4cm]{\allGraphsDir/EiffelBase_53/EiffelBase_53--#1.pdf} &
	\includegraphics[width=5.4cm]{\allGraphsDir/EiffelProgramAnalysis_74/EiffelProgramAnalysis_74--#1.pdf} &
	\includegraphics[width=5.4cm]{\allGraphsDir/GoboKernel_70/GoboKernel_70--#1.pdf} \\
	\includegraphics[width=5.4cm]{\allGraphsDir/GoboStructure_69/GoboStructure_69--#1.pdf} &
	\includegraphics[width=5.4cm]{\allGraphsDir/GoboTime_71/GoboTime_71--#1.pdf} &
	\includegraphics[width=5.4cm]{\allGraphsDir/GoboUtility_68/GoboUtility_68--#1.pdf} &
	\includegraphics[width=5.4cm]{\allGraphsDir/GoboXML_67/GoboXML_67--#1.pdf}
 \end{tabular}	
\end{center}
\caption{#2 (Eiffel projects)}
\end{figure}
\newpage
\begin{figure}[!htb]
\begin{center}
    \begin{tabular}{cccc}
	\includegraphics[width=5.4cm]{\allGraphsDir/Boogie_10/Boogie_10--#1.pdf} &
	\includegraphics[width=5.4cm]{\allGraphsDir/CCI_6/CCI_6--#1.pdf} &
	\includegraphics[width=5.4cm]{\allGraphsDir/Dafny_11/Dafny_11--#1.pdf} &
	\includegraphics[width=5.4cm]{\allGraphsDir/Labs_5/Labs_5--#1.pdf} \\
	\includegraphics[width=5.4cm]{\allGraphsDir/Quickgraph_18/Quickgraph_18--#1.pdf} &
	\includegraphics[width=5.4cm]{\allGraphsDir/Rxx_16/Rxx_16--#1.pdf} &
	\includegraphics[width=5.4cm]{\allGraphsDir/shweet_8/shweet_8--#1.pdf}
 \end{tabular}	
\end{center}
\caption{#2 (C\# projects)}
\end{figure}
\newpage
\begin{figure}[!htb]
\begin{center}
    \begin{tabular}{cccc}
	\includegraphics[width=5.4cm]{\allGraphsDir/DirectVCGen_205/DirectVCGen_205--#1.pdf} &
	\includegraphics[width=5.4cm]{\allGraphsDir/ESCJava_207/ESCJava_207--#1.pdf} &
	\includegraphics[width=5.4cm]{\allGraphsDir/JavaFE_202/JavaFE_202--#1.pdf} \\
	\includegraphics[width=5.4cm]{\allGraphsDir/Logging_203/Logging_203--#1.pdf} &
	\includegraphics[width=5.4cm]{\allGraphsDir/RCC_208/RCC_208--#1.pdf} &
	\includegraphics[width=5.4cm]{\allGraphsDir/Umbra_209/Umbra_209--#1.pdf}
 \end{tabular}	
\end{center}
\caption{#2 (Java projects)}
\end{figure}
}

\newcommand{\printFlat}[2]{
\begin{figure}[!htb]
\begin{center}
    \begin{tabular}{cccc}
	\includegraphics[width=5.4cm]{\allGraphsDir/AutoTest_72/AutoTest_72-#1.pdf} &
	\includegraphics[width=5.4cm]{\allGraphsDir/EiffelBase_53/EiffelBase_53-#1.pdf} &
	\includegraphics[width=5.4cm]{\allGraphsDir/EiffelProgramAnalysis_74/EiffelProgramAnalysis_74-#1.pdf} &
	\includegraphics[width=5.4cm]{\allGraphsDir/GoboKernel_70/GoboKernel_70-#1.pdf} \\
	\includegraphics[width=5.4cm]{\allGraphsDir/GoboStructure_69/GoboStructure_69-#1.pdf} &
	\includegraphics[width=5.4cm]{\allGraphsDir/GoboTime_71/GoboTime_71-#1.pdf} &
	\includegraphics[width=5.4cm]{\allGraphsDir/GoboUtility_68/GoboUtility_68-#1.pdf} &
	\includegraphics[width=5.4cm]{\allGraphsDir/GoboXML_67/GoboXML_67-#1.pdf}
 \end{tabular}	
\end{center}
\caption{#2 (Eiffel projects)}
\end{figure}
\newpage
\begin{figure}[!htb]
\begin{center}
    \begin{tabular}{cccc}
	\includegraphics[width=5.4cm]{\allGraphsDir/Boogie_10/Boogie_10-#1.pdf} &
	\includegraphics[width=5.4cm]{\allGraphsDir/CCI_6/CCI_6-#1.pdf} &
	\includegraphics[width=5.4cm]{\allGraphsDir/Dafny_11/Dafny_11-#1.pdf} &
	\includegraphics[width=5.4cm]{\allGraphsDir/Labs_5/Labs_5-#1.pdf} \\
	\includegraphics[width=5.4cm]{\allGraphsDir/Quickgraph_18/Quickgraph_18-#1.pdf} &
	\includegraphics[width=5.4cm]{\allGraphsDir/Rxx_16/Rxx_16-#1.pdf} &
	\includegraphics[width=5.4cm]{\allGraphsDir/shweet_8/shweet_8-#1.pdf}
 \end{tabular}	
\end{center}
\caption{#2 (C\# projects)}
\end{figure}
\newpage
\begin{figure}[!htb]
\begin{center}
    \begin{tabular}{cccc}
	\includegraphics[width=5.4cm]{\allGraphsDir/DirectVCGen_205/DirectVCGen_205-#1.pdf} &
	\includegraphics[width=5.4cm]{\allGraphsDir/ESCJava_207/ESCJava_207-#1.pdf} &
	\includegraphics[width=5.4cm]{\allGraphsDir/JavaFE_202/JavaFE_202-#1.pdf} \\
	\includegraphics[width=5.4cm]{\allGraphsDir/Logging_203/Logging_203-#1.pdf} &
	\includegraphics[width=5.4cm]{\allGraphsDir/RCC_208/RCC_208-#1.pdf} &
	\includegraphics[width=5.4cm]{\allGraphsDir/Umbra_209/Umbra_209-#1.pdf}
 \end{tabular}	
\end{center}
\caption{#2 (Java projects)}
\end{figure}
}

\section{Non-flat classes} 

The following pages show the plots for the 21 projects of our study, considering non-flat classes.
The dotted lines, when present, mark mean values of the various quantities.
The thin continuous lines that do not appear in the legend track the total number of routines, classes, or both, whose absolute values are scaled to the range determined by the main data represented in the graph.
When two of such thin lines are present, the red one tracks the number of classes and the aquamarine one tracks the number of routines.
When only one (red) thin line is present, it tracks the number of routines or classes according to the main content of the graph.

\begin{landscape}
\changepage{}{}{-10mm}{-10mm}{}{100mm}{}{}{}

\printNonFlat{change-types-nonempty}{Non flat: Changed routines (with nonempty unchanged spec)}
\newpage

\printNonFlat{change-types-spec}{Non-flat: Routines with changed specification (with or without body changing)}
\newpage

\printNonFlat{change-types}{Non-flat: changed routines (types of change)}
\newpage

\printNonFlat{classes-avg}{Non-flat: Average invariant strength and \% of classes with invariant}
\newpage

\printNonFlat{classes}{Non-flat: \# of classes, classes with invariants, invariant clauses}
\newpage

\printNonFlat{preORpost-boolean}{Non-flat: \# routines with specification}
\newpage

\printNonFlat{spec-boolean-percent}{Non-flat: \% routines and classes with specification}
\newpage

\printNonFlat{spec-boolean}{Non-flat: \# routines and classes with specification}
\newpage

\printNonFlat{spec-ratio}{Non-flat: Specification clauses per routine or class}
\newpage

\printNonFlat{spec-strength}{Non-flat: Average specification strength per routine or class}
\newpage

\printNonFlat{weak-strong}{Non-flat: Weakening and strengthening}
\newpage

\printNonFlat{public-vs-nonpublic}{Non-flat: \% routines with specification: public vs nonpublic}
\newpage

\printNonFlat{attr-inv}{Non-flat: Weakening and strengthening of class invariants}

\end{landscape}

\clearpage
\newpage

\section{Flat classes}

The following pages show the plots for the 21 projects of our study, considering flat classes.

\begin{landscape}

\printFlat{flat-change-types-nonempty}{Flat: Changed routines (with nonempty unchanged spec)}
\newpage

\printFlat{flat-change-types-spec}{Flat: Routines with changed specification (with or without body changing)}
\newpage

\printFlat{flat-change-types}{Flat: changed routines (types of change)}
\newpage

\printFlat{flat-classes-avg}{Flat: Average invariant strength and \% of classes with invariant}
\newpage

\printFlat{flat-classes}{Flat: \# of classes, classes with invariants, invariant clauses}
\newpage

\printFlat{flat-preORpost-boolean}{Flat: \# routines with specification}
\newpage

\printFlat{flat-spec-boolean-percent}{Flat: \% routines and classes with specification}
\newpage

\printFlat{flat-spec-boolean}{Flat: \# routines and classes with specification}
\newpage

\printFlat{flat-spec-ratio}{Flat: Specification clauses per routine or class}
\newpage

\printFlat{flat-spec-strength}{Flat: Average specification strength per routine or class}
\newpage

\printFlat{flat-weak-strong}{Flat: Weakening and strengthening}
\newpage

\printFlat{flat-public-vs-nonpublic}{Flat: \% routines with specification: public vs nonpublic}
\newpage

\printFlat{flat-attr-inv}{Flat: Weakening and strengthening of class invariants}

\end{landscape}

%% file: dose-data.tex
\changepage{}{}{10mm}{10mm}{}{10mm}{}{}{}

\section{Data About the Control Group} \label{sec:control-group-dose}

The following tables report the same measures as those used in the main
analysis, but target 10 student projects discussed at the end of
Section~\ref{sec:threatsToValidity}.

\input{\tablesLoc/dose_overalls-avg.tex}
\input{\tablesLoc/dose_overalls-avg-flat.tex}

\clearpage
\newpage
\input{\tablesLoc/dose_overall-nonflat.tex}
\input{\tablesLoc/dose_overall-flat.tex}

\clearpage
\newpage
\input{\tablesLoc/dose_change-major.tex}
\input{\tablesLoc/dose_change-sum.tex}

\clearpage
\newpage
\input{\tablesLoc/dose_correlate-flatnonflat.tex}

\clearpage
\newpage
\input{\tablesLoc/dose_correlate-spec.tex}
\input{\tablesLoc/dose_correlate-spec-flat.tex}

\clearpage
\newpage
\input{\tablesLoc/dose_correlate-privatepublic.tex}
\input{\tablesLoc/dose_correlate-privatepublic-flat.tex}

\clearpage
\newpage
\input{\tablesLoc/dose_contract-types.tex}


%% file: dose_overalls-avg.tex
\begin{sidewaystable}[ht]
\begin{center}
{\tiny
\begin{tabular}{|l|rrrr|rrrr|rrrr|rrrr|rrrr|rrrr|rrrr|rrrr|rrrr|}
  \hline
 &
\multicolumn{4}{c|}{\textsc{\# classes}} &
\multicolumn{4}{c|}{\textsc{\% classes inv}} &
\multicolumn{4}{c|}{\textsc{\# routines}} &
\multicolumn{4}{c|}{\textsc{\% routines spec}} &
\multicolumn{4}{c|}{\textsc{\% routines pre}} &
\multicolumn{4}{c|}{\textsc{\% routines post}} &
\multicolumn{4}{c|}{\textsc{avg routines pre}} &
\multicolumn{4}{c|}{\textsc{avg routines post}} &
\multicolumn{4}{c|}{\textsc{avg inv}} 
\\
\textbf{Project} & \mm & \mmedian & \MM & \std & \mm & \mmedian & \MM & \std & \mm & \mmedian & \MM & \std & \mm & \mmedian & \MM & \std & \mm & \mmedian & \MM & \std & \mm & \mmedian & \MM & \std & \mm & \mmedian & \MM & \std & \mm & \mmedian & \MM & \std & \mm & \mmedian & \MM & \std \\ 
  \hline
dose 76 & 0 & 16 & 28 & 6 & 0 & 0.53 & 1 & 0.12 & 0 & 34 & 86 & 26 & 0 & 0.36 & 1 & 0.18 & 0 & 0.29 & 1 & 0.15 & 0 & 0.23 & 1 & 0.18 & 0 & 1.46 & 2.12 & 0.23 & 0 & 2 & 3 & 0.57 & 0 & 2.88 & 3.12 & 0.61 \\ 
  dose 77 & 0 & 24 & 25 & 6 & 0 & 0.83 & 0.88 & 0.17 & 0 & 150 & 193 & 57 & 0 & 0.73 & 0.88 & 0.14 & 0 & 0.26 & 0.36 & 0.1 & 0 & 0.61 & 0.88 & 0.15 & 0 & 1.52 & 1.71 & 0.45 & 0 & 4.07 & 5.2 & 1.36 & 0 & 1.81 & 2.46 & 0.41 \\ 
  dose 78 & 0 & 33 & 35 & 8 & 0 & 0.37 & 0.41 & 0.08 & 0 & 93 & 96 & 25 & 0 & 0.12 & 0.16 & 0.02 & 0 & 0.06 & 0.12 & 0.02 & 0 & 0.1 & 0.12 & 0.04 & 0 & 1.67 & 1.8 & 0.32 & 0 & 1.22 & 1.56 & 0.51 & 0 & 1.92 & 1.92 & 0.4 \\ 
  dose 80 & 0 & 9 & 9 & 1 & 0 & 0.25 & 0.44 & 0.06 & 0 & 35 & 43 & 12 & 0 & 0.39 & 0.83 & 0.13 & 0 & 0.21 & 0.83 & 0.18 & 0 & 0.28 & 0.58 & 0.08 & 0 & 1.5 & 1.5 & 0.25 & 0 & 1.5 & 1.69 & 0.23 & 0 & 1.67 & 5 & 1.37 \\ 
  dose 86 & 0 & 37 & 79 & 22 & 0 & 0.28 & 1 & 0.19 & 0 & 172 & 406 & 107 & 0 & 0.49 & 0.68 & 0.15 & 0 & 0.41 & 0.56 & 0.12 & 0 & 0.24 & 0.3 & 0.08 & 0 & 1.4 & 1.64 & 0.19 & 0 & 1.55 & 2.44 & 0.44 & 0 & 1.57 & 3 & 0.44 \\
\hline
\end{tabular}
}
\caption{Specification overall statistics with non-flattened classes.}
\label{tab:overalls-avg}
\end{center}
\end{sidewaystable}

%% file: dose_overalls-avg-flat.tex
\begin{sidewaystable}[ht]
\begin{center}
{\tiny
\begin{tabular}{|l|rrrr|rrrr|rrrr|rrrr|rrrr|rrrr|rrrr|rrrr|rrrr|}
  \hline
 &
\multicolumn{4}{c|}{\textsc{\# classes}} &
\multicolumn{4}{c|}{\textsc{\% classes inv}} &
\multicolumn{4}{c|}{\textsc{\# routines}} &
\multicolumn{4}{c|}{\textsc{\% routines spec}} &
\multicolumn{4}{c|}{\textsc{\% routines pre}} &
\multicolumn{4}{c|}{\textsc{\% routines post}} &
\multicolumn{4}{c|}{\textsc{avg routines pre}} &
\multicolumn{4}{c|}{\textsc{avg routines post}} &
\multicolumn{4}{c|}{\textsc{avg inv}} 
\\
\textbf{Project} & \mm & \mmedian & \MM & \std & \mm & \mmedian & \MM & \std & \mm & \mmedian & \MM & \std & \mm & \mmedian & \MM & \std & \mm & \mmedian & \MM & \std & \mm & \mmedian & \MM & \std & \mm & \mmedian & \MM & \std & \mm & \mmedian & \MM & \std & \mm & \mmedian & \MM & \std \\ 
  \hline
dose 76 & 0 & 16 & 28 & 6 & 0 & 0.79 & 1 & 0.14 & 0 & 46 & 122 & 37 & 0 & 0.39 & 1 & 0.13 & 0 & 0.33 & 1 & 0.11 & 0 & 0.2 & 1 & 0.16 & 0 & 1.29 & 1.91 & 0.21 & 0 & 1.71 & 2.5 & 0.46 & 0 & 2.33 & 2.67 & 0.45 \\ 
  dose 77 & 0 & 24 & 25 & 6 & 0 & 0.88 & 0.93 & 0.16 & 0 & 184 & 228 & 65 & 0 & 0.69 & 0.91 & 0.15 & 0 & 0.25 & 0.34 & 0.1 & 0 & 0.57 & 0.91 & 0.17 & 0 & 1.42 & 1.62 & 0.42 & 0 & 3.71 & 4.78 & 1.27 & 0 & 2 & 2.58 & 0.43 \\ 
  dose 78 & 0 & 33 & 35 & 8 & 0 & 0.69 & 0.72 & 0.2 & 0 & 156 & 162 & 43 & 0 & 0.31 & 0.34 & 0.09 & 0 & 0.25 & 0.27 & 0.07 & 0 & 0.26 & 0.28 & 0.11 & 0 & 2 & 2.1 & 0.47 & 0 & 1.41 & 1.5 & 0.55 & 0 & 2.52 & 2.54 & 0.63 \\ 
  dose 80 & 0 & 9 & 9 & 1 & 0 & 0.25 & 0.44 & 0.06 & 0 & 35 & 44 & 12 & 0 & 0.38 & 0.83 & 0.13 & 0 & 0.21 & 0.83 & 0.18 & 0 & 0.28 & 0.58 & 0.08 & 0 & 1.5 & 1.5 & 0.25 & 0 & 1.5 & 1.69 & 0.23 & 0 & 1.67 & 5 & 1.37 \\ 
  dose 86 & 0 & 37 & 79 & 22 & 0 & 0.56 & 1 & 0.19 & 0 & 206 & 574 & 169 & 0 & 0.49 & 0.6 & 0.11 & 0 & 0.31 & 0.51 & 0.1 & 0 & 0.29 & 0.35 & 0.1 & 0 & 1.44 & 1.7 & 0.2 & 0 & 1.32 & 2.44 & 0.47 & 0 & 1.46 & 3 & 0.44 \\
\hline
\end{tabular}
}
\caption{Specification overall statistics with flattened classes.}
\label{tab:overalls-avg-flat}
\end{center}
\end{sidewaystable}

%% file: dose_overall-nonflat.tex
\begin{sidewaystable}[ht]
\begin{center}
{\scriptsize
\begin{tabular}{|l|lrr|rrrr|rrrr|rrrr|rrrr|rrrr|rrrr|rrrr|}
  \hline
 &
 &  &  &
\multicolumn{4}{c|}{\textsc{LOC}} &
\multicolumn{4}{c|}{\textsc{\# classes}} &
\multicolumn{4}{c|}{\textsc{\% classes inv}} &
\multicolumn{4}{c|}{\textsc{\# routines}} &
\multicolumn{4}{c|}{\textsc{\% routines spec}} &
\multicolumn{4}{c|}{\textsc{\% routines pre}} &
\multicolumn{4}{c|}{\textsc{\% routines post}} 
\\
\textbf{Project} & \textbf{lang} & \textbf{rev} & \textbf{age} & \mm & \mmedian & \MM & \std & \mm & \mmedian & \MM & \std & \mm & \mmedian & \MM & \std & \mm & \mmedian & \MM & \std & \mm & \mmedian & \MM & \std & \mm & \mmedian & \MM & \std & \mm & \mmedian & \MM & \std \\ 
  \hline
dose 76 & Eiffel & 151 & 8 & 0 & 2051 & 3494 & 1055 & 0 & 16 & 28 & 6 & 0 & 0.53 & 1 & 0.12 & 0 & 34 & 86 & 26 & 0 & 0.36 & 1 & 0.18 & 0 & 0.29 & 1 & 0.15 & 0 & 0.23 & 1 & 0.18 \\ 
  dose 77 & Eiffel & 70 & 8 & 0 & 3486 & 4349 & 1177 & 0 & 24 & 25 & 6 & 0 & 0.83 & 0.88 & 0.17 & 0 & 150 & 193 & 57 & 0 & 0.73 & 0.88 & 0.14 & 0 & 0.26 & 0.36 & 0.1 & 0 & 0.61 & 0.88 & 0.15 \\ 
  dose 78 & Eiffel & 73 & 8 & 0 & 2908 & 3208 & 930 & 0 & 33 & 35 & 8 & 0 & 0.37 & 0.41 & 0.08 & 0 & 93 & 96 & 25 & 0 & 0.12 & 0.16 & 0.02 & 0 & 0.06 & 0.12 & 0.02 & 0 & 0.1 & 0.12 & 0.04 \\ 
  dose 80 & Eiffel & 143 & 7 & 0 & 1556 & 1944 & 516 & 0 & 9 & 9 & 1 & 0 & 0.25 & 0.44 & 0.06 & 0 & 35 & 43 & 12 & 0 & 0.39 & 0.83 & 0.13 & 0 & 0.21 & 0.83 & 0.18 & 0 & 0.28 & 0.58 & 0.08 \\ 
  dose 86 & Eiffel & 170 & 8 & 0 & 3086 & 8827 & 2519 & 0 & 37 & 79 & 22 & 0 & 0.28 & 1 & 0.19 & 0 & 172 & 406 & 107 & 0 & 0.49 & 0.68 & 0.15 & 0 & 0.41 & 0.56 & 0.12 & 0 & 0.24 & 0.3 & 0.08 \\ 
  dose 87 & Eiffel & 235 & 7 & 0 & 2358 & 4460 & 1287 & 0 & 14 & 18 & 4 & 0 & 0.2 & 0.67 & 0.11 & 0 & 134 & 163 & 46 & 0 & 0.49 & 0.84 & 0.17 & 0 & 0.37 & 0.59 & 0.1 & 0 & 0.22 & 0.59 & 0.13 \\
\hline
\end{tabular}
}
\caption{Project specification statistics for non-flattened classes.}
\label{tab:overall-nonflat}
\end{center}
\end{sidewaystable}

%% file: dose_overall-flat.tex
\begin{sidewaystable}[ht]
\begin{center}
{\scriptsize
\begin{tabular}{|l|lrr|rrrr|rrrr|rrrr|rrrr|rrrr|rrrr|rrrr|}
  \hline
 &
 &  &  &
\multicolumn{4}{c|}{\textsc{LOC}} &
\multicolumn{4}{c|}{\textsc{\# classes}} &
\multicolumn{4}{c|}{\textsc{\% classes inv}} &
\multicolumn{4}{c|}{\textsc{\# routines}} &
\multicolumn{4}{c|}{\textsc{\% routines spec}} &
\multicolumn{4}{c|}{\textsc{\% routines pre}} &
\multicolumn{4}{c|}{\textsc{\% routines post}} 
\\
\textbf{Project} & \textbf{lang} & \textbf{rev} & \textbf{age} & \mm & \mmedian & \MM & \std & \mm & \mmedian & \MM & \std & \mm & \mmedian & \MM & \std & \mm & \mmedian & \MM & \std & \mm & \mmedian & \MM & \std & \mm & \mmedian & \MM & \std & \mm & \mmedian & \MM & \std \\ 
  \hline
dose 76 & Eiffel & 151 & 8 & 0 & 2051 & 3494 & 1055 & 0 & 16 & 28 & 6 & 0 & 0.79 & 1 & 0.14 & 0 & 46 & 122 & 37 & 0 & 0.39 & 1 & 0.13 & 0 & 0.33 & 1 & 0.11 & 0 & 0.2 & 1 & 0.16 \\ 
  dose 77 & Eiffel & 70 & 8 & 0 & 3486 & 4349 & 1177 & 0 & 24 & 25 & 6 & 0 & 0.88 & 0.93 & 0.16 & 0 & 184 & 228 & 65 & 0 & 0.69 & 0.91 & 0.15 & 0 & 0.25 & 0.34 & 0.1 & 0 & 0.57 & 0.91 & 0.17 \\ 
  dose 78 & Eiffel & 73 & 8 & 0 & 2908 & 3208 & 930 & 0 & 33 & 35 & 8 & 0 & 0.69 & 0.72 & 0.2 & 0 & 156 & 162 & 43 & 0 & 0.31 & 0.34 & 0.09 & 0 & 0.25 & 0.27 & 0.07 & 0 & 0.26 & 0.28 & 0.11 \\ 
  dose 80 & Eiffel & 143 & 7 & 0 & 1556 & 1944 & 516 & 0 & 9 & 9 & 1 & 0 & 0.25 & 0.44 & 0.06 & 0 & 35 & 44 & 12 & 0 & 0.38 & 0.83 & 0.13 & 0 & 0.21 & 0.83 & 0.18 & 0 & 0.28 & 0.58 & 0.08 \\ 
  dose 86 & Eiffel & 170 & 8 & 0 & 3086 & 8827 & 2519 & 0 & 37 & 79 & 22 & 0 & 0.56 & 1 & 0.19 & 0 & 206 & 574 & 169 & 0 & 0.49 & 0.6 & 0.11 & 0 & 0.31 & 0.51 & 0.1 & 0 & 0.29 & 0.35 & 0.1 \\ 
  dose 87 & Eiffel & 235 & 7 & 0 & 2358 & 4460 & 1287 & 0 & 14 & 18 & 4 & 0 & 0.2 & 0.67 & 0.11 & 0 & 134 & 163 & 46 & 0 & 0.49 & 0.84 & 0.17 & 0 & 0.37 & 0.59 & 0.1 & 0 & 0.22 & 0.59 & 0.13 \\
\hline
\end{tabular}
}
\caption{Project specification statistics for flattened classes.}
\label{tab:overall-flat}
\end{center}
\end{sidewaystable}

%% file: dose_change-major.tex
\begin{table*}[ht]
\begin{center}
{\scriptsize
\begin{tabular}{|c|rrrrrrrrrrrr|}
  \hline
&
\begin{sideways} chang vs unch spec \end{sideways} &
\begin{sideways} NE chang vs unch spec \end{sideways} &
\begin{sideways} pre weak vs strong \end{sideways} &
\begin{sideways} NE pre weak vs strong \end{sideways} &
\begin{sideways} post weak vs strong \end{sideways} &
\begin{sideways} NE post weak vs strong \end{sideways} &
\begin{sideways} inv weak vs strong \end{sideways} &
\begin{sideways} NE inv weak vs strong \end{sideways} &
\begin{sideways} $+$A inv weak vs same \end{sideways} &
\begin{sideways} $+$A inv strong vs same \end{sideways} &
\begin{sideways} $-$A inv weak vs same \end{sideways} &
\begin{sideways} $-$A inv strong vs same \end{sideways} \\

  \hline
$V$ & 0.00E+00 & 3.05E+01 & 1.00E+00 & 3.50E+00 & 2.50E+00 & 0.00E+00 & 1.50E+00 & 2.00E+00 & 0.00E+00 & 0.00E+00 & 0.00E+00 & 0.00E+00 \\ 
  $p$ & 1.95E-03 & 3.74E-01 & 1.26E-02 & 1.40E-02 & 1.18E-02 &  & 4.18E-02 & 8.90E-02 & 5.86E-03 & 5.86E-03 & 5.83E-03 & 5.76E-03 \\ 
  $\Delta(\mu)$ & -4.10E+01 & 2.50E+00 & -3.50E+00 & -1.00E+00 & -3.00E+00 & 0.00E+00 & -2.00E+00 & -5.00E-01 & -1.50E+01 & -1.40E+01 & -4.50E+00 & -4.50E+00 \\ 
  $d$ & -1.19E+00 & 3.93E-01 & -7.05E-01 & -4.39E-01 & -9.94E-01 & 0.00E+00 & -7.04E-01 & -4.66E-01 & -1.41E+00 & -1.33E+00 & -1.14E+00 & -1.19E+00 \\ 
   \hline
$V$ & 0.00E+00 & 3.05E+01 & 1.00E+00 & 3.50E+00 & 2.50E+00 & 0.00E+00 & 1.50E+00 & 2.00E+00 & 0.00E+00 & 0.00E+00 & 0.00E+00 & 0.00E+00 \\ 
  $p$ & 1.95E-03 & 3.74E-01 & 1.26E-02 & 1.40E-02 & 1.18E-02 &  & 4.18E-02 & 8.90E-02 & 5.86E-03 & 5.86E-03 & 5.83E-03 & 5.76E-03 \\ 
  $\Delta(\mu)$ & -4.10E+01 & 2.50E+00 & -3.50E+00 & -1.00E+00 & -3.00E+00 & 0.00E+00 & -2.00E+00 & -5.00E-01 & -1.50E+01 & -1.40E+01 & -4.50E+00 & -4.50E+00 \\ 
  $d$ & -1.19E+00 & 3.93E-01 & -7.05E-01 & -4.39E-01 & -9.94E-01 & 0.00E+00 & -7.04E-01 & -4.66E-01 & -1.41E+00 & -1.33E+00 & -1.14E+00 & -1.19E+00 \\ 
   \hline
\end{tabular}
}
\caption{Change analysis with majority for non-flattened (top) and flattened (bottom) classes.}
\label{tab:change-major}
\end{center}
\end{table*}

%% file: dose_change-sum.tex
\begin{table*}[ht]
\begin{center}
{\scriptsize
\begin{tabular}{|c|rrrrrrrrrrrr|}
  \hline
&
\begin{sideways} chang vs unch spec \end{sideways} &
\begin{sideways} NE chang vs unch spec \end{sideways} &
\begin{sideways} pre weak vs strong \end{sideways} &
\begin{sideways} NE pre weak vs strong \end{sideways} &
\begin{sideways} post weak vs strong \end{sideways} &
\begin{sideways} NE post weak vs strong \end{sideways} &
\begin{sideways} inv weak vs strong \end{sideways} &
\begin{sideways} NE inv weak vs strong \end{sideways} &
\begin{sideways} $+$A inv weak vs same \end{sideways} &
\begin{sideways} $+$A inv strong vs same \end{sideways} &
\begin{sideways} $-$A inv weak vs same \end{sideways} &
\begin{sideways} $-$A inv strong vs same \end{sideways} \\

  \hline
$V$ & 0.00E+00 & 2.80E+01 & 6.00E+00 & 0.00E+00 & 0.00E+00 & 0.00E+00 & 5.00E+00 & 2.50E+00 & 0.00E+00 & 0.00E+00 & 0.00E+00 & 0.00E+00 \\ 
  $p$ & 1.95E-03 & 5.54E-01 & 3.22E-02 & 5.67E-03 & 5.89E-03 &  & 4.36E-02 & 3.32E-02 & 1.95E-03 & 5.89E-03 & 5.86E-03 & 5.86E-03 \\ 
  $\Delta(\mu)$ & -1.31E+02 & 2.50E+00 & -8.00E+00 & -2.00E+00 & -8.00E+00 & 0.00E+00 & -3.00E+00 & -1.00E+00 & -2.00E+01 & -1.70E+01 & -7.00E+00 & -8.00E+00 \\ 
  $d$ & -9.99E-01 & 2.61E-01 & -7.41E-01 & -5.63E-01 & -9.48E-01 & 0.00E+00 & -8.55E-01 & -5.94E-01 & -1.34E+00 & -1.28E+00 & -1.26E+00 & -1.29E+00 \\ 
   \hline
$V$ & 0.00E+00 & 2.80E+01 & 6.00E+00 & 0.00E+00 & 0.00E+00 & 0.00E+00 & 5.00E+00 & 2.50E+00 & 0.00E+00 & 0.00E+00 & 0.00E+00 & 0.00E+00 \\ 
  $p$ & 1.95E-03 & 5.54E-01 & 3.22E-02 & 5.67E-03 & 5.89E-03 &  & 4.36E-02 & 3.32E-02 & 1.95E-03 & 5.89E-03 & 5.86E-03 & 5.86E-03 \\ 
  $\Delta(\mu)$ & -1.31E+02 & 2.50E+00 & -8.00E+00 & -2.00E+00 & -8.00E+00 & 0.00E+00 & -3.00E+00 & -1.00E+00 & -2.00E+01 & -1.70E+01 & -7.00E+00 & -8.00E+00 \\ 
  $d$ & -9.99E-01 & 2.61E-01 & -7.41E-01 & -5.63E-01 & -9.48E-01 & 0.00E+00 & -8.55E-01 & -5.94E-01 & -1.34E+00 & -1.28E+00 & -1.26E+00 & -1.29E+00 \\ 
   \hline
\end{tabular}
}
\caption{Change analysis for non-flattened (top) and flattened (bottom) classes.}
\label{tab:change-sum}
\end{center}
\end{table*}

%% file: dose_correlate-flatnonflat.tex
\begin{sidewaystable}[ht]
\begin{center}
{\footnotesize
\begin{tabular}{|l|rr|rr|rr|rr|rr|}
  \hline
 &
\multicolumn{2}{|c}{\textsc{\% routines w/ spec}} &
\multicolumn{2}{|c}{\textsc{\% classes w/ inv}} &
\multicolumn{2}{|c}{\textsc{avg pre strength}} &
\multicolumn{2}{|c}{\textsc{avg post strength}} &
\multicolumn{2}{|c|}{\textsc{avg inv strength}}
\\
\textbf{Project} & $p$ & $\tau$ & $p$ & $\tau$ & $p$ & $\tau$ & $p$ & $\tau$ & $p$ & $\tau$ \\ 
  \hline
dose 76 & 0.00E+00 & 0.67 & 0.00E+00 & 0.89 & 0.00E+00 & 0.76 & 0.00E+00 & 0.95 & 0.00E+00 & 0.66 \\ 
  dose 77 & 0.00E+00 & 0.96 & 0.00E+00 & 0.86 & 0.00E+00 & 0.81 & 0.00E+00 & 0.94 & 0.00E+00 & 0.79 \\ 
  dose 78 & 1.05E-09 & 0.53 & 7.58E-08 & 0.48 & 4.44E-16 & 0.78 & 0.00E+00 & 0.80 & 0.00E+00 & 0.86 \\ 
  dose 80 & 0.00E+00 & 0.94 & 0.00E+00 & 1.00 & 0.00E+00 & 1.00 & 0.00E+00 & 1.00 & 0.00E+00 & 1.00 \\ 
  dose 86 & 0.00E+00 & 0.80 & 7.07E-11 & 0.35 & 0.00E+00 & 0.68 & 0.00E+00 & 0.87 & 3.42E-05 & 0.23 \\ 
  dose 87 & 0.00E+00 & 1.00 & 0.00E+00 & 1.00 & 0.00E+00 & 1.00 & 0.00E+00 & 1.00 & 0.00E+00 & 0.85 \\ 
  dose 88 & 0.00E+00 & 0.93 & 0.00E+00 & 0.89 & 0.00E+00 & 0.89 & 0.00E+00 & 0.83 & 6.66E-16 & 0.60 \\ 
  dose 89 & 0.00E+00 & 0.58 & 0.00E+00 & 1.00 & 0.00E+00 & 0.76 & 0.00E+00 & 0.77 & 0.00E+00 & 1.00 \\ 
  dose 90 & 0.00E+00 & 0.71 & 0.00E+00 & 1.00 & 0.00E+00 & 0.76 & 3.15E-10 & 0.38 & 0.00E+00 & 1.00 \\ 
  dose 92 & 0.00E+00 & 0.71 & 0.00E+00 & 1.00 & 9.70E-04 & 0.14 & 0.00E+00 & 0.41 & 0.00E+00 & 1.00 \\ 
   \hline
\textbf{Overall} & 0.00E+00 & 0.59 & 0.00E+00 & 0.87 & 0.00E+00 & 0.62 & 0.00E+00 & 0.87 & 0.00E+00 & 0.79 \\ 
   \hline
\end{tabular}
}
\caption{Correlation of measures in flat and non-flat classes.}
\label{tab:correlate-flatnonflat}
\end{center}
\end{sidewaystable}

%% file: dose_correlate-spec.tex
\begin{table*}[ht]
\begin{center}
{\scriptsize
\begin{tabular}{|l|rr|rr|rr|rr|rr|rr|rr|}
  \hline
 &
\multicolumn{2}{|c}{\textsc{\% spec/\# routines}} &
\multicolumn{2}{|c}{\textsc{\# spec/\# routines}} &
\multicolumn{2}{|c}{\textsc{\% inv/\# classes}} &
\multicolumn{2}{|c}{\textsc{\# inv/\# classes}} &
\multicolumn{2}{|c}{\textsc{avg/\% pre}} &
\multicolumn{2}{|c}{\textsc{avg/\% post}} &
\multicolumn{2}{|c|}{\textsc{avg/\% inv}}
\\
\textbf{Project} & $p$ & $\tau$ & $p$ & $\tau$ & $p$ & $\tau$ & $p$ & $\tau$ & $p$ & $\tau$ & $p$ & $\tau$ & $p$ & $\tau$ \\ 
  \hline
dose 76 & 8.31E-31 & -0.66 & 0.00E+00 & 0.65 & 8.79E-45 & -0.85 & 0.00E+00 & 0.89 & 1.11E-15 & 0.47 & 3.20E-09 & 0.34 & 9.84E-01 & -0.00 \\ 
  dose 77 & 7.21E-08 & -0.45 & 0.00E+00 & 0.94 & 1.77E-07 & 0.49 & 0.00E+00 & 0.91 & 5.04E-03 & 0.24 & 1.27E-03 & -0.27 & 9.03E-09 & 0.52 \\ 
  dose 78 & 6.71E-01 & -0.04 & 2.82E-04 & 0.34 & 5.70E-04 & 0.31 & 0.00E+00 & 0.82 & 2.90E-10 & -0.61 & 1.39E-04 & 0.35 & 6.82E-14 & 0.71 \\ 
  dose 80 & 4.91E-20 & -0.57 & 0.00E+00 & 0.56 & 3.38E-04 & -0.28 & 9.36E-01 & 0.01 & 2.40E-25 & -0.72 & 1.30E-04 & 0.24 & 9.15E-01 & -0.01 \\ 
  dose 86 & 1.61E-08 & -0.30 & 0.00E+00 & 0.79 & 1.67E-15 & -0.43 & 0.00E+00 & 0.90 & 0.00E+00 & 0.65 & 2.49E-06 & 0.25 & 1.72E-13 & -0.41 \\ 
  dose 87 & 2.06E-06 & -0.21 & 0.00E+00 & 0.51 & 6.42E-24 & -0.50 & 3.93E-05 & 0.22 & 1.93E-02 & -0.11 & 1.28E-08 & -0.26 & 0.00E+00 & 0.66 \\
\hline
\end{tabular}
}
\caption{Correlation of measures of specification in non-flat classes.}
\label{tab:correlate-spec}
\end{center}
\end{table*}

%% file: dose_correlate-spec-flat.tex
\begin{table*}[ht]
\begin{center}
{\scriptsize
\begin{tabular}{|l|rr|rr|rr|rr|rr|rr|rr|}
  \hline
 &
\multicolumn{2}{|c}{\textsc{\% spec/\# routines}} &
\multicolumn{2}{|c}{\textsc{\# spec/\# routines}} &
\multicolumn{2}{|c}{\textsc{\% inv/\# classes}} &
\multicolumn{2}{|c}{\textsc{\# inv/\# classes}} &
\multicolumn{2}{|c}{\textsc{avg/\% pre}} &
\multicolumn{2}{|c}{\textsc{avg/\% post}} &
\multicolumn{2}{|c|}{\textsc{avg/\% inv}}
\\
\textbf{Project} & $p$ & $\tau$ & $p$ & $\tau$ & $p$ & $\tau$ & $p$ & $\tau$ & $p$ & $\tau$ & $p$ & $\tau$ & $p$ & $\tau$ \\ 
  \hline
dose 76 & 6.87E-15 & -0.45 & 0.00E+00 & 0.84 & 9.71E-42 & -0.82 & 0.00E+00 & 0.94 & 0.00E+00 & 0.71 & 5.77E-09 & 0.34 & 3.70E-09 & -0.35 \\ 
  dose 77 & 9.81E-08 & -0.44 & 0.00E+00 & 0.93 & 3.10E-04 & 0.34 & 0.00E+00 & 0.91 & 8.46E-07 & 0.41 & 9.37E-04 & -0.28 & 8.18E-09 & 0.54 \\ 
  dose 78 & 9.22E-07 & 0.42 & 0.00E+00 & 0.84 & 5.35E-03 & 0.25 & 0.00E+00 & 0.94 & 1.68E-03 & 0.27 & 1.79E-06 & 0.41 & 2.17E-04 & 0.34 \\ 
  dose 80 & 7.68E-25 & -0.63 & 0.00E+00 & 0.54 & 3.38E-04 & -0.28 & 9.36E-01 & 0.01 & 8.74E-25 & -0.70 & 5.63E-05 & 0.25 & 9.15E-01 & -0.01 \\ 
  dose 86 & 6.16E-02 & -0.10 & 0.00E+00 & 0.96 & 2.04E-05 & 0.23 & 0.00E+00 & 0.94 & 0.00E+00 & 0.51 & 3.84E-04 & -0.19 & 4.42E-02 & 0.11 \\ 
  dose 87 & 2.06E-06 & -0.21 & 0.00E+00 & 0.51 & 6.42E-24 & -0.50 & 3.93E-05 & 0.22 & 1.93E-02 & -0.11 & 1.28E-08 & -0.26 & 0.00E+00 & 0.70 \\
\hline
\end{tabular}
}
\caption{Correlation of measures of specification in flat classes.}
\label{tab:correlate-spec-flat}
\end{center}
\end{table*}

%% file: dose_correlate-privatepublic.tex
\begin{table*}[ht]
\begin{center}
{\footnotesize
\begin{tabular}{|l|rr|rr|rr|}
  \hline
 &
\multicolumn{2}{|c}{\textsc{\% pre public/private}} &
\multicolumn{2}{|c}{\textsc{\% post public/private}} &
\multicolumn{2}{|c|}{\textsc{\% spec public/private}}
\\
\textbf{Project} & $p$ & $\tau$ & $p$ & $\tau$ & $p$ & $\tau$ \\ 
  \hline
dose 76 & 0.00E+00 & 0.51 & 0.00E+00 & 0.83 & 0.00E+00 & 0.48 \\ 
  dose 77 & 8.76E-10 & 0.53 & 1.93E-14 & 0.67 & 5.99E-08 & 0.47 \\ 
  dose 78 & 3.01E-12 & -0.62 & 3.85E-03 & 0.25 & 2.07E-01 & 0.11 \\ 
  dose 80 & 3.86E-01 & 0.05 & 2.25E-06 & -0.29 & 3.13E-01 & -0.06 \\ 
  dose 86 & 0.00E+00 & 0.72 & 0.00E+00 & 0.62 & 0.00E+00 & 0.69 \\ 
  dose 87 & 6.68E-02 & -0.10 & 6.69E-01 & -0.02 & 3.17E-01 & 0.05 \\ 
  dose 88 & 5.89E-02 & 0.13 & 2.09E-10 & 0.42 & 8.45E-02 & 0.12 \\ 
  dose 89 & 2.39E-04 & 0.17 & 0.00E+00 & 0.45 & 1.55E-06 & 0.23 \\ 
  dose 90 & 1.79E-01 & 0.09 & 1.11E-01 & -0.10 & 5.93E-01 & -0.03 \\ 
  dose 92 & 4.66E-15 & 0.33 & 3.62E-03 & -0.12 & 2.75E-01 & 0.05 \\ 
   \hline
\textbf{Overall} & 0.00E+00 & 0.23 & 0.00E+00 & 0.21 & 0.00E+00 & 0.16 \\ 
   \hline
\end{tabular}
}
\caption{Correlation of measures between public and private routines in non-flat classes.}
\label{tab:correlate-privatepublic}
\end{center}
\end{table*}

%% file: dose_correlate-privatepublic-flat.tex
\begin{table*}[ht]
\begin{center}
{\footnotesize
\begin{tabular}{|l|rr|rr|rr|}
  \hline
 &
\multicolumn{2}{|c}{\textsc{\% pre public/private}} &
\multicolumn{2}{|c}{\textsc{\% post public/private}} &
\multicolumn{2}{|c|}{\textsc{\% spec public/private}}
\\
\textbf{Project} & $p$ & $\tau$ & $p$ & $\tau$ & $p$ & $\tau$ \\ 
  \hline
dose 76 & 7.26E-01 & -0.02 & 0.00E+00 & 0.79 & 2.30E-09 & 0.34 \\ 
  dose 77 & 1.11E-13 & 0.64 & 4.22E-15 & 0.68 & 1.76E-10 & 0.55 \\ 
  dose 78 & 1.84E-01 & 0.11 & 4.17E-03 & 0.25 & 2.43E-01 & 0.10 \\ 
  dose 80 & 2.09E-01 & 0.08 & 6.54E-05 & -0.24 & 8.02E-01 & 0.02 \\ 
  dose 86 & 0.00E+00 & 0.64 & 0.00E+00 & 0.49 & 0.00E+00 & 0.66 \\ 
  dose 87 & 6.68E-02 & -0.10 & 4.40E-01 & 0.04 & 1.49E-02 & 0.12 \\ 
  dose 88 & 7.62E-01 & -0.02 & 7.97E-11 & 0.43 & 3.13E-01 & 0.07 \\ 
  dose 89 & 2.50E-01 & -0.05 & 0.00E+00 & 0.44 & 1.41E-01 & 0.07 \\ 
  dose 90 & 8.95E-01 & 0.01 & 3.20E-05 & -0.27 & 1.71E-06 & -0.31 \\ 
  dose 92 & 0.00E+00 & 0.44 & 0.00E+00 & 0.66 & 0.00E+00 & 0.79 \\ 
   \hline
\textbf{Overall} & 0.00E+00 & 0.18 & 0.00E+00 & 0.20 & 1.05E-11 & 0.12 \\ 
   \hline
\end{tabular}
}
\caption{Correlation of measures between public and private routines in flat classes.}
\label{tab:correlate-privatepublic-flat}
\end{center}
\end{table*}

%% file: dose_contract-types.tex
\begin{sidewaystable}[ht]
\begin{center}
{\scriptsize
%
}
\caption{Syntactic elements found in contracts (non-flattened classes).}
\label{tab:contract-types}
\end{center}
\end{sidewaystable}

%% file: main.bbl
\begin{thebibliography}{10}
\providecommand{\url}[1]{\texttt{#1}}
\providecommand{\urlprefix}{URL }

\bibitem{AraujoBriandLabiche11}
Araujo, W., Briand, L.C., Labiche, Y.: On the effectiveness of contracts as
  test oracles in the detection and diagnosis of race conditions and deadlocks
  in concurrent object-oriented software. In: ESEM. pp. 10--19 (2011)

\bibitem{DBLP:journals/cacm/BarnettFLMSV11}
Barnett, M., F{\"a}hndrich, M., Leino, K.R.M., M{\"u}ller, P., Schulte, W.,
  Venter, H.: Specification and verification: the {S}pec\# experience. Comm.
  ACM  54(6),  81--91 (2011)

\bibitem{BoyapatiKM02}
Boyapati, C., Khurshid, S., Marinov, D.: {K}orat: automated testing based on
  {J}ava predicates. In: ISSTA. pp. 123--133 (2002)

\bibitem{BrunHolmesEtAl2011}
Brun, Y., Holmes, R., Ernst, M.D., Notkin, D.: Proactive detection of
  collaboration conflicts. In: ESEC/FSE. pp. 168--178 (2011)

\bibitem{Chalin06}
Chalin, P.: Are practitioners writing contracts? In: The RODIN Book. LNCS, vol.
  4157, p. 100 (2006)

\bibitem{CheonL02}
Cheon, Y., Leavens, G.T.: A simple and practical approach to unit testing: The
  {JML} and {JUnit} way. In: ECOOP. pp. 231--255 (2002)

\bibitem{COAT-data}
\url{http://se.inf.ethz.ch/data/coat/}

\bibitem{codecontract-webpage}
\url{http://research.microsoft.com/en-us/projects/contracts/}

\bibitem{CurinoMZ08}
Curino, C., Moon, H.J., Zaniolo, C.: Graceful database schema evolution: the
  prism workbench. PVLDB  1(1),  761--772 (2008)

\bibitem{DietlDEMS11}
Dietl, W., Dietzel, S., Ernst, M.D., Muslu, K., Schiller, T.W.: Building and
  using pluggable type-checkers. In: ICSE. pp. 681--690. ACM (2011)

\bibitem{DrossopoulouFMS08}
Drossopoulou, S., Francalanza, A., M{\"u}ller, P., Summers, A.J.: A unified
  framework for verification techniques for object invariants. In: ECOOP 2008
  -- Object-Oriented Programming. pp. 412--437. Lecture Notes in Computer
  Science, Springer (2008)

\bibitem{ErnstPGMPTX07}
Ernst, M.D., Perkins, J.H., Guo, P.J., McCamant, S., Pacheco, C., Tschantz,
  M.S., Xiao, C.: The {D}aikon system for dynamic detection of likely
  invariants. Sci. Comput. Program.  69,  35--45 (2007)

\bibitem{HowSpecChange-TR-17082012}
Estler, H.C., Furia, C.A., Nordio, M., Piccioni, M., Meyer, B.: Contracts in
  practice. \url{http://arxiv.org/abs/1211.4775} (2013), extended version with
  appendix

\bibitem{FahndrichBL10}
F{\"a}hndrich, M., Barnett, M., Logozzo, F.: Embedded contract languages. In:
  SAC. pp. 2103--2110. ACM (2010)

\bibitem{FluriWuerschGall07}
Fluri, B., W{\"u}rsch, M., Gall, H.: Do code and comments co-evolve? on the
  relation between source code and comment changes. In: WCRE. pp. 70--79. IEEE
  (2007)

\bibitem{Duque09}
Garc\'{i}a-Duque, J., Pazos-Arias, J., L\'{o}pez-Nores, M.,
  Blanco-Fern\'{a}ndez, Y., Fern\'{a}ndez-Vilas, A., D\'{i}az-Redondo, R.,
  Ramos-Cabrer, M., Gil-Solla, A.: Methodologies to evolve formal
  specifications through refinement and retrenchment in an analysis-revision
  cycle. Requirements Engineering  14,  129--153 (2009)

\bibitem{Gaudel07}
Gaudel, M.C.: Software testing based on formal specification. In: Testing
  Techniques in Software Engineering. Lecture Notes in Computer Science, vol.
  6153, pp. 215--242. Springer (2007)

\bibitem{GermanCubranicStorey05}
Germ{\'a}n, D.M., Cubranic, D., Storey, M.A.D.: A framework for describing and
  understanding mining tools in software development. In: MSR (2005)

\bibitem{HarmanKLMY10}
Harman, M., Kim, S.G., Lakhotia, K., McMinn, P., Yoo, S.: Optimizing for the
  number of tests generated in search based test data generation with an
  application to the oracle cost problem. In: ICST Workshops. pp. 182--191
  (2010)

\bibitem{HenkelRD07}
Henkel, J., Reichenbach, C., Diwan, A.: Discovering documentation for {J}ava
  container classes. IEEE Trans. Software Eng.  33(8),  526--543 (2007)

\bibitem{HieronsBBCDDGHKKLSVWZ09}
Hierons, R.M., Bogdanov, K., Bowen, J.P., Cleaveland, R., Derrick, J., Dick,
  J., Gheorghe, M., Harman, M., Kapoor, K., Krause, P., L{\"u}ttgen, G.,
  Simons, A.J.H., Vilkomir, S.A., Woodward, M.R., Zedan, H.: Using formal
  specifications to support testing. ACM Comput. Surv.  41(2) (2009)

\bibitem{HindleBirdZimmermannNagappan12}
Hindle, A., Bird, C., Zimmermann, T., Nagappan, N.: Relating requirements to
  implementation via topic analysis. In: ICSM (2012)

\bibitem{JiangHSZX08}
Jiang, Y., Hou, S.S., Shan, J., Zhang, L., Xie, B.: An approach to testing
  black-box components using contract-based mutation. International Journal of
  Software Engineering and Knowledge Engineering  18(1),  93--117 (2008)

\bibitem{jmlspec-webpage}
\url{http://www.jmlspecs.org}

\bibitem{DBLP:conf/icse/KimCK11}
Kim, M., Cai, D., Kim, S.: An empirical investigation into the role of
  {API}-level refactorings during software evolution. In: ICSE. pp. 151--160.
  ACM (2011)

\bibitem{kind-webpage}
\url{http://kindsoftware.com}

\bibitem{KiniryZ08}
Kiniry, J.R., Zimmerman, D.M.: Secret ninja formal methods. In: FM. LNCS, vol.
  5014, pp. 214--228. Springer (2008)

\bibitem{KudrjavetsNB06}
Kudrjavets, G., Nagappan, N., Ball, T.: Assessing the relationship between
  software assertions and faults: An empirical investigation. In: ISSRE. pp.
  204--212 (2006)

\bibitem{JML99}
Leavens, G.T., Baker, A.L., Ruby, C.: {JML}: A notation for detailed design.
  In: Behavioral Specifications of Businesses and Systems, pp. 175--188. Kluwer
  Academic Publishers (1999)

\bibitem{MarinovK01}
Marinov, D., Khurshid, S.: {TestEra}: A novel framework for automated testing
  of {J}ava programs. In: ASE. p.~22 (2001)

\bibitem{Martin96smallsample}
Martin, J.K., Hirschberg, D.S.: Small sample statistics for classification
  error rates {II}. Tech. rep., CS Department, UC Irvine (1996),
  \url{http://goo.gl/Ec8oD}

\bibitem{Meyer97}
Meyer, B.: Object Oriented Software Construction. Prentice Hall PTR, 2 edn.
  (1997)

\bibitem{Meyer2009}
Meyer, B., Fiva, A., Ciupa, I., Leitner, A., Wei, Y., Stapf, E.: Programs that
  test themselves. Computer  42(9),  46--55 (2009)

\bibitem{bertrand-void-safe-eiffel}
Meyer, B., Kogtenkov, A., Stapf, E.: Avoid a {V}oid: the eradication of null
  dereferencing. In: Reflections on the Work of C.A.R. Hoare, pp. 189--211.
  Springer (2010)

\bibitem{MonkS93}
Monk, S.R., Sommerville, I.: Schema evolution in {OODBs} using class
  versioning. SIGMOD Record  22(3),  16--22 (1993)

\bibitem{MullerTH02}
M{\"u}ller, M.M., Typke, R., Hagner, O.: Two controlled experiments concerning
  the usefulness of assertions as a means for programming. In: ICSM. pp. 84--92
  (2002)

\bibitem{NeamtiuFosterHicks05}
Neamtiu, I., Foster, J.S., Hicks, M.W.: Understanding source code evolution
  using abstract syntax tree matching. In: MSR (2005)

\bibitem{NEFM-TR-20110919}
Nordio, M., Estler, H.C., Furia, C.A., Meyer, B.: Collaborative software
  development on the web. \url{http://arxiv.org/abs/1105.0768} (September 2011)

\bibitem{openjml-webpage}
\url{http://openjml.org}

\bibitem{Parkinson2007}
Parkinson, M.: Class invariants: The end of the road? (position paper) (2007)

\bibitem{DBLP:journals/cacm/Parnas72a}
Parnas, D.L.: On the criteria to be used in decomposing systems into modules.
  Commun. ACM  15(12),  1053--1058 (1972)

\bibitem{Parnas-FOSE}
Parnas, D.L.: Precise documentation: The key to better software. In: The Future
  of Software Engineering. pp. 125--148. Springer (2011)

\bibitem{PiccioniOriolMeyer12}
Piccioni, M., Oriol, M., Meyer, B.: Class schema evolution for persistent
  object-oriented software: Model, empirical study, and automated support. IEEE
  TSE  (2012), to appear

\bibitem{PolikarpovaCM09}
Polikarpova, N., Ciupa, I., Meyer, B.: A comparative study of
  programmer-written and automatically inferred contracts. In: ISSTA. pp.
  93--104 (2009)

\bibitem{PFPWM-ICSE13}
Polikarpova, N., Furia, C.A., Pei, Y., Wei, Y., Meyer, B.: What good are strong
  specifications? In: ICSE. pp. 257--266. ACM (2013)

\bibitem{semicola}
Polikarpova, N., Tschannen, J., Furia, C.A., Meyer, B.: Flexible invariants
  through semantic collaboration. In: FM. LNCS, Springer (2014)

\bibitem{PradelG13}
Pradel, M., Gross, T.R.: Automatic testing of sequential and concurrent
  substitutability. In: ICSE. pp. 282--291. ACM (2013)

\bibitem{DBLP:journals/smr/RiccaPTTC07}
Ricca, F., Penta, M.D., Torchiano, M., Tonella, P., Ceccato, M.: How design
  notations affect the comprehension of {W}eb applications. Journal of Software
  Maintenance  19(5),  339--359 (2007)

\bibitem{writingcontracts}
Schiller, T.W., Donohue, K., Coward, F., Ernst, M.D.: Writing and enforcing
  contract specifications. In: ICSE. ACM (2014)

\bibitem{StaatsWH11}
Staats, M., Whalen, M.W., Heimdahl, M.P.E.: Programs, tests, and oracles. In:
  ICSE. pp. 391--400 (2011)

\bibitem{TemperoYN13}
Tempero, E.D., Yang, H.Y., Noble, J.: What programmers do with inheritance in
  {J}ava. In: ECOOP. Lecture Notes in Computer Science, vol. 7920, pp.
  577--601. Springer (2013)

\bibitem{TillmannH08}
Tillmann, N., de~Halleux, J.: {Pex}--{W}hite box test generation for {.NET}.
  In: TAP. pp. 134--153 (2008)

\bibitem{WasylkowskiZ11}
Wasylkowski, A., Zeller, A.: Mining temporal specifications from object usage.
  Autom. Softw. Eng.  18(3-4),  263--292 (2011)

\bibitem{WeiFKM11}
Wei, Y., Furia, C.A., Kazmin, N., Meyer, B.: Inferring better contracts. In:
  ICSE. pp. 191--200 (2011)

\bibitem{WeiETAL11}
Wei, Y., Roth, H., Furia, C.A., Pei, Y., Horton, A., Steindorfer, M., Nordio,
  M., Meyer, B.: Stateful testing: Finding more errors in code and contracts.
  In: ASE. IEEE (2011)

\bibitem{Woodcock-survey}
Woodcock, J., Larsen, P.G., Bicarregui, J., Fitzgerald, J.: Formal methods:
  Practice and experience. ACM CSUR  41(4) (2009)

\bibitem{Xie06}
Xie, T.: Augmenting automatically generated unit-test suites with regression
  oracle checking. In: ECOOP. Lecture Notes in Computer Science, vol. 4067, pp.
  380--403. Springer (2006)

\bibitem{ZaidmanVanEtAl2008}
Zaidman, A., Van~Rompaey, B., Demeyer, S., van Deursen, A.: Mining software
  repositories to study co-evolution of production and test code. In: ICST. pp.
  220 --229 (2008)

\bibitem{ZimmermanN10}
Zimmerman, D.M., Nagmoti, R.: {JMLUnit}: The next generation. In: FoVeOOS.
  LNCS, vol. 6528, pp. 183--197. Springer (2010)

\end{thebibliography}
